\documentclass[11pt]{article}
\usepackage[margin=1in]{geometry}
\IfFileExists{lmodern.sty}{\usepackage[T1]{fontenc}\usepackage{lmodern}}{}
\usepackage{microtype}
\usepackage{amsmath,amssymb,amsthm,mathtools,bm}
\usepackage{graphicx,booktabs,array,float,placeins}
\usepackage[dvipsnames,table]{xcolor}
\usepackage{enumitem}
\usepackage[numbers,sort&compress]{natbib}
\usepackage[colorlinks=true,linkcolor=blue!55!black,citecolor=blue!60!black,urlcolor=blue!65!black]{hyperref}

\newcommand{\Prb}{\mathbb P}\newcommand{\E}{\mathbb E}\newcommand{\I}{\mathrm I}\newcommand{\Hh}{\mathrm H}
\newcommand{\yh}{\hat y}
\newtheorem{proposition}{Proposition}
\theoremstyle{remark}
\setlist{itemsep=2pt,topsep=3pt}

\title{\bfseries How Path Integral Diffusion Decides a Finite Spin Glass:\\[2pt]
\large Stochastic Localization, Level-Resolved Information and Spectral Decision Times}
\author{Michael (Misha) Chertkov\thanks{Program in Applied Mathematics and Department of Mathematics, University of Arizona, Tucson,
AZ 85721, USA (chertkov@arizona.edu; ORCID 0000-0002-6758-515X).}}
\date{\today}

\begin{document}\maketitle

\begin{abstract}
When an Artificial Intelligence (AI) diffusion-type generator samples a frustrated spin glass, which macroscopic alternatives does it decide first, and when? We study this
for bridge-diffusion models of the Path Integral Diffusion (PID) type started from a point source. We show that every isotropic PID is a time-changed Gaussian stochastic localization, and conversely every signal-to-noise schedule is realized by a PID protocol. What a continuous-time generator decides therefore does not depend on the protocol, which
only sets when and at what cost. We describe the generator's state by its \emph{partial commitments}: the probabilities it assigns, at each
moment, to nested classes of configurations fixed by the target. The partial commitments are martingales that move only along class contrasts. Their ensemble
average gives the information about each level at an exact rate set by squared magnetization contrasts.  In the balanced, effectively binary approximation, a split reaches half of its class information near $\gamma=1.044/\lambda_P$, where $\lambda_P$ is the leading eigenvalue of the parent covariance. The theory is applied to Bethe-lattice and Sherrington--Kirkpatrick (SK) glasses with up to $24$ spins. It is observed generically that  a trajectory first shifts its weight to one of the two mirror halves of the zero-field measure and then to the glassy classes inside it. These classes appear near the glass transition, keep their identity on cooling and are absent in a ferromagnet on the same graphs. Sparse (Bethe) and dense (SK) glasses differ in how often and how strongly such classes exist, not in how trajectories commit to
them. Nested glassy levels have similar $\lambda_P$ and are decided together, so a trajectory commits in one or two distinct events---the mirror choice and a compound glassy decision---which separate further with coupling and size, as the spectra predict. For a generator that takes
finitely many steps, the protocol matters through the late, local stage rather than the macroscopic decisions. Sampled posteriors at $N=32,64$
and small trained denoisers show the same sequence.
\end{abstract}

\section{Introduction}\label{sec:intro}

A generator that turns noise into a sample of a structured distribution must, at some point, commit to one of the distribution's
macroscopic alternatives: a class of images, a molecular conformation, a symmetry-broken sector. Which alternatives are decided first, and
when? For AI diffusion models this question has been studied as \emph{speciation}---trajectories commit to a class at a time set by the spectrum of
the data covariance, and later, in a \emph{collapse} or memorization regime, to individual samples
\citep{biroli2023generative,biroli2024dynamical,behjoo2025uturn,raya2023spontaneous,ambrogioni2025statistical}. Successive commitments follow
class structure \citep{achilli2026theory,sclocchi2025phase,sclocchi2025probing}, and transitions along the denoising path can make sampling hard
\citep{ghio2024sampling}.

\paragraph{Related speciation studies.} After the initial arXiv submission of this paper, the author learned, through correspondence with M.~M\'ezard, of the contemporaneous analysis by Garnier-Brun, Biroli, M\'ezard and Bouchaud~\citep{GarnierBrun2026Foam}. In an analytically tractable high-dimensional, long-range correlated random landscape, they obtain a \emph{speciation cascade} that resolves progressively finer branches of a full replica-symmetry-breaking hierarchy and a \emph{foam} of sharp interfaces in the score potential. Li and Chen~\citep{LiChen2024Critical} previously derived non-asymptotic bounds on feature-dependent \emph{critical windows} for mixtures of strongly log-concave distributions. The present finite-size work gives exact partial-commitment and information-production identities for any fixed operational hierarchy, together with approximate decision-time predictions.

\paragraph{Why spin glasses.} Our interest is in AI generation in general---images, molecules, materials, physical fields
\citep{yang2023diffusion,noe2020mlmolsim,debreuck2025crystal,cranmer2023lattice}; Sec.~\ref{sec:discussion} returns to these applications. For such data neither the classes nor the generator's posterior along its path can be computed, so what a generator decides, and when, can only be inferred.
We therefore pose the question where everything can be computed exactly and the answer is least trivial. Spin glasses qualify on both counts: their Gibbs measure is known exactly. Frustration organizes the low-temperature configurations into many competing groups---in the thermodynamic limit into the nested hierarchy of Parisi's solution \citep{parisi1979infinite,parisi1983order,mezard1987sgtb,panchenko2013ultrametricity}---so
the target is as hierarchical as any distribution we know. And spin glasses are among the hardest distributions to sample
\citep{barahona1982computational,elalaoui2022sampling}. We use the two canonical mean-field glasses, both with a glass transition at finite
temperature: the Sherrington--Kirkpatrick (SK) model, with couplings between all pairs of spins \citep{sherrington1975solvable}, and the
Bethe-lattice glass, with couplings on a sparse random graph \citep{viana1985phase,mezard2001bethe}. At $N\le24$ the phase transitions of these models are smoothed into crossovers, and all quantities of interest---classes, posteriors and
information---can be computed without approximation by enumerating all $2^N$ configurations. The tools we develop---classes defined from the target, the generator's
partial commitments to them, and a spectral prediction for when each is decided---require only a posterior or a trained denoiser, and so apply to other targets as well.

\begin{figure}[!htbp]\centering
\includegraphics[width=\linewidth]{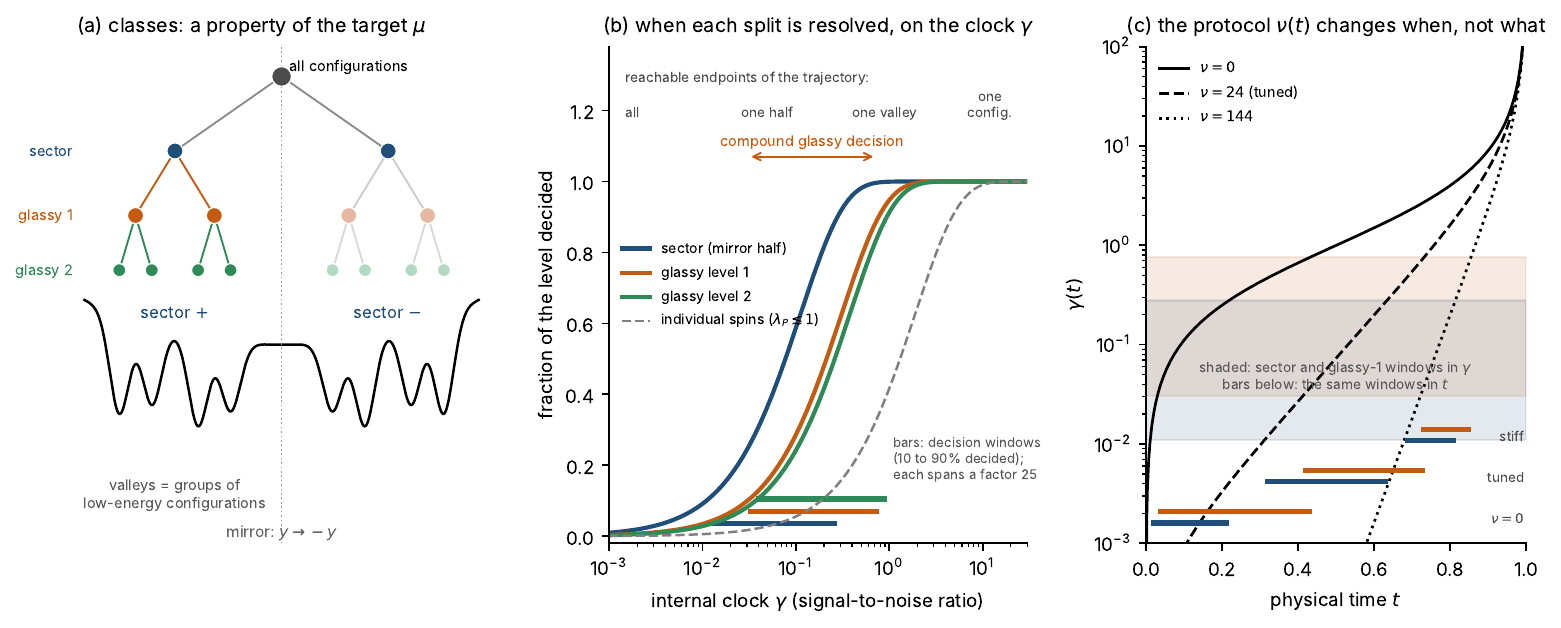}
\caption{\textbf{Classes, partial commitments and clocks.} (a) Classes are a property of the target: the two mirror halves of the zero-field
measure (sectors), and inside each half nested groups of low-energy configurations (glassy classes) in different valleys of the energy; the right
tree mirrors the left one. (b) Every split divides a \emph{parent} class into two \emph{children}: the whole measure into
the two sectors, a sector into its two glassy classes, and so on. On the internal clock $\gamma$, the information that the generator holds about which child contains its endpoint is compared with the balanced binary-channel benchmark $I_{\rm bin}(\gamma\lambda_P)$, where $\lambda_P$ is the leading
eigenvalue of the parent's correlation matrix (Sec.~\ref{sec:channel}); a large $\lambda_P$ means an early decision. The curves are drawn with the
values of the $N=24$ glass of Fig.~\ref{fig:anatomy}: $\lambda_P=13.5$ for the sector (whose parent is the whole measure), and $\lambda_P=4.9$ and
$4.0$ for the first two glassy levels. Bars mark the benchmark $10$--$90\%$ windows, each a factor of about $25$ wide. Glassy levels with similar $\lambda_P$ are decided together, as one compound decision. Individual
spins follow at $\gamma\sim1$. (c) The protocol $\nu(t)$ (Eq.~\eqref{eq:SOT-objective}) only maps the internal clock onto physical time: the same
windows in $\gamma$ (shaded) fall at different times $t$ (bars) for three harmonic stiffnesses.}
\label{fig:schematic}
\end{figure}

\paragraph{Three notions.} We study one \emph{finite} Ising spin glass in zero field and one generative AI generator -- Path Integral Diffusion (PID) from a point
source \citep{behjoo2025hpid,chertkov2026adapid,arxiv2605_02961}. Three notions organize the paper (Fig.~\ref{fig:schematic}).
\begin{itemize}[leftmargin=*]
\item \emph{Classes} are groups of configurations fixed by the target: the two mirror halves of the measure, and inside each half nested groups of
low-energy configurations in different valleys of the energy (panel a).

\item \emph{Partial commitments} are the generator's state: at every moment a trajectory assigns a probability to every class. These probabilities start at the equilibrium class masses and end at one for the classes of the endpoint. A trajectory has \emph{committed} to a class
when its probability is close to one. Following the partial commitments shows how a trajectory descends the tree, level by level or several levels at once (panel b).

\item The \emph{clock}. PID has a physical time $t$ and a protocol, the stiffness $\nu(t)$ of a harmonic confinement. What the generator knows depends only on an internal clock $\gamma$, the signal-to-noise ratio; the protocol only maps $\gamma$ onto $t$ (panel c).
\end{itemize}

\paragraph{Results -- Theory.}
\begin{itemize}
	\item  (i) PID from a point source and Gaussian stochastic localization are one process seen on different clocks: each isotropic PID
protocol is a time change of it, and each signal-to-noise schedule comes from some protocol (Proposition~\ref{prop:sl}). The generator's knowledge is the same glass in a random field correlated with the endpoint, the
tilt of stochastic localization \citep{eldan2013thin,elalaoui2022sampling,montanari2023sampling}. 

\item (ii) The partial commitments to the target's classes evolve as martingales driven only by class contrasts, and averaging them over
trajectories gives an exact rate for the information about each level (Proposition~\ref{prop:levelrate}).

\item (iii) In an approximately rigid and balanced binary split, the generator resolves a one-bit Gaussian channel with half-information time $\gamma_{1/2}\simeq1.044/\lambda_P$ and a $10$--$90\%$ window spanning a factor about $25$ in $\gamma$. Unequal child masses require a mass-aware channel; at fixed split contrast their half-information time need not follow the balanced formula.
\citet{biroli2024dynamical} gave the inverse leading eigenvalue for the first speciation. We derive the constant and the window, apply the estimate
recursively over a hierarchy, and measure its accuracy.
\end{itemize}

\paragraph{Results -- Experiments.} The experiments use exactly enumerated Bethe and Sherrington--Kirkpatrick (SK) glasses with up to $24$ spins, sampled posteriors at $N=32,64$, matched controls and trained denoisers.
\begin{itemize}
	\item 
(iv) Every trajectory first commits to one of the two mirror halves---a choice every ordered zero-field Ising model makes---and then to glassy
classes within it. These classes are found near and below the glass transition, persist as the target is cooled, and do not exist in a
ferromagnet on the same graphs (Secs.~\ref{sec:anatomy}--\ref{sec:temperature}). 

\item (v) Nested glassy levels are decided together: a trajectory commits $2.6$ times on average at $N=16$ but in $1.5$ distinct events, the
mirror choice and a compound glassy decision, which separate further with coupling and size as the spectra predict. Bethe and SK glasses give
the same per-trajectory picture; they differ in how often glassy classes exist and how strong they are. 

\item (vi) With an exact drift and a finite number of steps, the choice of protocol is felt mainly in the final, local stage, where individual
spins are fixed, rather than in the macroscopic decisions; this gives a practical rule for allocating steps (Sec.~\ref{sec:protocol}).

\end{itemize}

\paragraph{Scope.} Everything is at finite $N$. \emph{Hierarchy} means a nested family of classes of one Gibbs measure whose splits pass stated
tests, including collectivity. Whether its levels are decided in sequence is measured, not assumed. We make no claim about replica-symmetry
breaking or $N\to\infty$. The representations used here presuppose the endpoint. The representations analyse what an ideal or learned generator decides and do
not make sampling easier.

\paragraph{Organization.} Section~\ref{sec:pid}: PID as stochastic localization and what the protocol can change. Section~\ref{sec:decisions}:
classes, partial commitments and when each split is decided. Section~\ref{sec:levels}: how classes are identified. Section~\ref{sec:experiments}:
experiments. Section~\ref{sec:discussion}: discussion. Appendices: proofs, a positive control with a known tree, contrast identities used as
numerical tools, numerical methods, and a map from figures to scripts. Code, data and notebooks are available at
\url{https://github.com/mchertkov/SpinGlassHierarchy}.

\FloatBarrier
\section{Path Integral Diffusion as time-changed stochastic localization}\label{sec:pid}

\subsection{Target and generator}
The target is one Ising instance in zero field on a graph $\mathcal G=(\mathcal V,\mathcal E)$, with vertices $\mathcal V=\{1,\dots,N\}$ (the spins)
and edges $\mathcal E\subseteq\{(i,j):1\le i<j\le N\}$ (the interacting pairs, each listed once). With couplings $J=(J_{ij}\mid(i,j)\in\mathcal E)$,
\begin{equation}\label{eq:target}
\mu(y)=\frac1{Z}\exp\Big(\sum_{(i,j)\in\mathcal E}J_{ij}y_iy_j\Big),\qquad y\in\{\pm1\}^N,\qquad J=\beta_JJ^{(0)} .
\end{equation}
Here $\beta_J$ is the inverse temperature of the target and $J^{(0)}=(J^{(0)}_{ij}\mid(i,j)\in\mathcal E)$ a fixed disorder realization of
Gaussian couplings. The experiments use two graphs (Sec.~\ref{sec:procedure}): a random three-regular graph, on which every spin has three
neighbours and $|\mathcal E|=3N/2$ (Bethe glass), and the complete graph, on which every pair interacts and $|\mathcal E|=N(N-1)/2$ (SK glass). The generator is the controlled diffusion
in $\mathbb R^N$
\begin{equation}\label{eq:sde}
dX_t=u(t,X_t)\,dt+dW_t,\qquad X_0=0,
\end{equation}
whose drift minimizes
\begin{equation}\label{eq:SOT-objective}
\E\int_0^1\big(\tfrac12\|u\|^2+\tfrac12\nu(t)\|X_t\|^2\big)dt\quad\text{subject to}\quad X_1\sim\mu .
\end{equation}
The \emph{protocol} $\nu(t)$, the stiffness of a harmonic confinement, is a property of the generator, not of the target. Eq.~(\ref{eq:SOT-objective}) is the harmonic PID introduced and discussed extensively in  \citep{behjoo2025hpid,chertkov2026adapid} and its $\nu=0$ case corresponds to the Schr\"odinger--F\"ollmer process \citep{follmer1985entropy,leonard2014survey,chen2021liaisons}.

Because $X_0=0$, the optimal path law for~\eqref{eq:SOT-objective} is the
Brownian reference law weighted by the harmonic confinement and an
endpoint-dependent factor~\citep{leonard2014survey,behjoo2025hpid,chertkov2026adapid}.
Conditioned on $Y=X_1=y$, the endpoint factor cancels and the process is
a Gaussian bridge of the reference law. Its drift is
$\kappa(t)y-c(t)x$, where the deterministic coefficients are fixed by
the confinement: $c'=c^2-\nu(t)$, $\kappa'=c\kappa$, with
$c(t),\kappa(t)\sim(1-t)^{-1}$ as $t\uparrow1$
(Appendix~\ref{app:proofs}). The optimal drift is the average of this
conditional bridge drift over the posterior law of $Y$, as made explicit
in~\eqref{eq:drift} below.

The conditional process is linear, with Gaussian one-time marginals
\begin{equation}\label{eq:bridge}
X_t\mid Y=y\sim\mathcal N\big(a(t)y,\sigma^2(t)I\big),\qquad
 a(t)=\frac{\gamma(t)}{\kappa(t)},\quad
 \sigma^2(t)=\frac{\gamma(t)}{\kappa^2(t)},\quad
 \gamma(t)=\int_0^t\kappa^2(s)\,ds.
\end{equation}
Here $a(t)$ is the bridge mean coefficient and $\sigma^2(t)$ its
coordinate variance. Consequently, $\gamma(t)=a^2(t)/\sigma^2(t)$ is
the signal-to-noise ratio for observing the endpoint through $X_t$. Its role as a protocol-independent clock is established in the next subsection. The derivation and the explicit constant-$\nu$ example are given in Appendix~\ref{app:proofs}.

\subsection{The generator's knowledge is the same glass in a singleton field}
What does the generator know about its endpoint at time $t$? By Bayes' rule, the posterior of the endpoint given $X_t=x$ is the prior $\mu(y)$ times
the Gaussian likelihood \eqref{eq:bridge}, $\exp\big(-\|x-a y\|^2/2\sigma^2\big)$. Because every spin configuration has the same length,
$\|y\|^2=N$, the likelihood depends on $y$ only through $\exp(\kappa\,x\cdot y)$, and
\begin{equation}\label{eq:posterior}
P_t(y\mid x)=\frac{1}{Z(h)}\exp\Big(\sum_{(i,j)\in\mathcal E}J_{ij}y_iy_j+h\cdot y\Big),\qquad h=\kappa(t)\,x .
\end{equation}
The posterior is the same Ising model as the target, with the same couplings, in an additional external field $h=\kappa x$ that acts on each
spin separately (a ``singleton'' field); the field carries what the generator has learned about its endpoint. This is the tilt of stochastic localization \citep{eldan2013thin,eldan2020taming}; for
the SK model \citet{elalaoui2022sampling} write the same Ising model in an external field.
The density of the generator is a Gaussian envelope
times $Z(\kappa x)/Z(0)$. Averaging the endpoint-conditioned bridge drift $\kappa(t)y-c(t)x$ over this posterior gives the optimal drift:
\begin{equation}\label{eq:drift}
\yh(t;x)=\E[Y\mid X_t=x]=\nabla_h\log Z(h)\big|_{h=\kappa x},\qquad u(t,x)=\kappa(t)\,\yh(t;x)-c(t)\,x .
\end{equation}
Generating a glass therefore amounts to computing, at every step, the magnetization $\yh$ of the same glass in a random field $h$. Exact enumeration,
Monte Carlo, belief propagation or a trained network are different ways of doing so; in this paper we use exact enumeration for $N\le24$.

\subsection{Time change, its converse, and what the protocol can change} \label{sec:clock}

\paragraph{From physical time to an internal clock.} So far everything has been written in the physical time $t\in[0,1]$, and every protocol
$\nu(t)$ gives a different process $X_t$, with a different drift and a different speed. We now change the time variable and find that, seen on a
suitable clock, all these processes are one and the same. The clue is in the posterior \eqref{eq:posterior}: the generator's knowledge of its
endpoint enters only through the field $h_t=\kappa(t)X_t$. Given the endpoint $y$, the bridge \eqref{eq:bridge} gives $X_t\sim\mathcal N(a y,\sigma^2I)$,
and therefore
\begin{equation}\label{eq:field-snr}
	h_t\mid y\;\sim\;\mathcal N\big(\kappa a\,y,\ \kappa^2\sigma^2 I\big)=\mathcal N\big(\gamma\,y,\ \gamma\,I\big),
\end{equation}
because $\kappa a=\gamma$ and $\kappa^2\sigma^2=a^2/\sigma^2=\gamma$. At every time, the field is a noisy observation of the endpoint, $\gamma y$
plus Gaussian noise of variance $\gamma$; how informative it is depends on $t$ and on the protocol only through the signal-to-noise ratio
$\gamma(t)$. This suggests using $\gamma$ itself as the clock, $t\to\gamma(t)$. Proposition~\ref{prop:sl} below shows that the statement holds not only at
each time separately but for whole trajectories: on the clock $\gamma$, the field process of every isotropic PID is the same universal process,
$H_\gamma=\gamma Y+B_\gamma$---a Brownian motion that drifts with velocity equal to the endpoint---which is the process of Gaussian stochastic
localization. Different protocols merely run through this process at different speeds in physical time. Conversely, every way of running through it
corresponds to some PID protocol.

\begin{proposition}[PID and stochastic localization]\label{prop:sl}
Let $H_\gamma=\gamma Y+B_\gamma$ with $Y\sim\mu$ and $B$ a standard Brownian motion independent of $Y$.
\begin{enumerate}[label=(\roman*),leftmargin=*]
\item For every unit-noise PID with isotropic confinement $\tfrac12\nu(t)\|x\|^2$, $\gamma'=\kappa^2$ and the field process satisfies
$h_t=\kappa(t)X_t=H_{\gamma(t)}$ in law, jointly in $t$ and $Y$. In the generator's own filtration $dH_\gamma=\yh\,d\gamma+d\bar B_\gamma$ with an
innovation Brownian motion $\bar B$.
\item Conversely, let $\gamma\in C^2[0,1)$ be increasing with $\gamma(0)=0$ and $(1-t)^2\gamma'(t)\to1$ as $t\to1$. Set $\kappa=\sqrt{\gamma'}$,
$c=\kappa'/\kappa$ and $\nu=c^2-c'$. Then $X_t=H_{\gamma(t)}/\kappa(t)$ solves $dX_t=[\kappa\yh-cX_t]\,dt+dW_t$ with $X_0=0$ and $X_1=Y$: it is the
optimal PID for the confinement $\tfrac12\nu(t)\|x\|^2$, which may be negative.
\item For an anisotropic confinement with matrix $B(t)$, the bridge is $\mathcal N(A_ty,\Sigma_t)$ and the tilt is
$\exp(y^\top A_t^\top\Sigma_t^{-1}x-\tfrac12y^\top\Gamma_ty)$ with the matrix clock $\Gamma_t=A_t^\top\Sigma_t^{-1}A_t$; its off-diagonal part adds
time-dependent pair couplings to \eqref{eq:posterior}. Matrix-valued localization schemes \citep{eldan2013thin,montanari2023sampling} are of this
type.
\end{enumerate}
\end{proposition}

Part (i) for $\nu=0$, and the analogous statement for Ornstein--Uhlenbeck diffusions, are the change of time and scale of
\citet[Sec.~1.3]{montanari2023sampling}; that losses and posteriors depend only on the signal-to-noise ratio underlies variational diffusion
models \citep{kingma2021variational}, and \citet{grenioux2024stochastic} use general schedules with MCMC posteriors along the path. What the
proposition adds is the explicit map between schedules and PID protocols, in both directions.

\paragraph{What the protocol cannot change.} By part (i) the generator's knowledge, as a function of $\gamma$, has the same law for every
protocol. Everything defined from it---the partial commitments of Sec.~\ref{sec:partial}, hence which classes are decided, in which order and with which
probabilities, how often a trajectory recrosses, how much information it holds---is therefore independent of $\nu(t)$. We work on the internal clock $\gamma$ throughout, and on the scaled clock
$\tau=N\gamma$ for macroscopic decisions.

\paragraph{What the protocol does change.}(a) \emph{When} each decision happens in physical time: a decision at $\gamma_*$ happens at
$t_*=\gamma^{-1}(\gamma_*)$. Because $\gamma(t)$ is increasing, the protocol can stretch or compress the intervals between decisions but never
reorder them. The explicit constant-confinement example is given in Appendix~\ref{app:proofs}. (b) \emph{What it costs}:

\begin{proposition}[Control cost and information]\label{prop:energy}
For every unit-noise isotropic PID and $0<t<1$,
\begin{equation}\label{eq:energy}
\frac12\int_0^t\E\|u(s,X_s)\|^2ds=\frac N2\int_0^t\big[a'(s)^2+c(s)^2\sigma^2(s)\big]ds-\I(Y;X_t).
\end{equation}
\end{proposition}
The first integral on the right is the mean drift energy of the
endpoint-conditioned bridges. The mutual information $\I(Y;X_t)$
is exactly the reduction in drift energy obtained by averaging
those bridge drifts over the posterior of $Y$ (Appendix~\ref{app:proofs}). For $\nu=0$ the baseline is $-\tfrac N2\log(1-t)$; since $\I(Y;X_t)\le N\log2$, the cost of reaching a discrete target
diverges as $t\to1$, so \eqref{eq:SOT-objective} is to be read as the limit of costs up to $t<1$. (c) \emph{How well a discretized generator works.} A generator that takes
finitely many steps in $t$ places them on the $\gamma$ axis according to its protocol; Sec.~\ref{sec:protocol} measures what this changes.

\FloatBarrier
\section{Classes, partial commitments and decision times}\label{sec:decisions}
This section sets up the language used in all experiments. It answers four questions in turn: \emph{what} can be decided (classes, which are a
property of the target, Sec.~\ref{sec:defs}); \emph{how} one trajectory decides (its partial commitments, Sec.~\ref{sec:partial}); \emph{what
the ensemble of trajectories learns} about each level (information per level, Sec.~\ref{sec:levelrate}); and \emph{when} each split is decided
(the binary-channel prediction, Sec.~\ref{sec:channel}).

\subsection{Classes: a property of the target}\label{sec:defs}
\paragraph{Classes, levels and hierarchies.} A \emph{class} is a set of spin configurations. A \emph{level} is a partition of the whole
configuration space $\{\pm1\}^N$ into classes: every configuration belongs to exactly one class of the level. A \emph{hierarchy} is a nested sequence
of levels $C_0\prec C_1\prec\dots\prec C_L$, in which every class of level $\ell$ lies inside one class of level $\ell-1$, called its \emph{parent};
the classes of level $\ell$ that lie inside a given class of level $\ell-1$ are its \emph{children}. In this paper every split is binary: a class
has either two children or none. We write $c\in C_\ell$ for a class of level $\ell$ and $\pi(c)\in C_{\ell-1}$ for its parent. When we speak of
``a class $c$'' without naming a level, $c$ may belong to any level, not only to the finest one: masses, magnetizations and partial commitments
are defined for every class of every level. Because each level is a partition of the whole space, a class that is not split further reappears
unchanged at the next level, as its own single child; in the tree diagrams of Figs.~\ref{fig:anatomy} and \ref{fig:temperature} such a class is
drawn as an open circle.

\paragraph{The branch of a configuration.} Every configuration $y$ lies in exactly one class of each level. We write $c_\ell(y)\in C_\ell$ for the class
of level $\ell$ that contains $y$. The classes containing $y$ form a nested chain,
\begin{equation}\label{eq:branch}
c_0(y)\supseteq c_1(y)\supseteq\dots\supseteq c_L(y),
\end{equation}
which we call the \emph{branch} of $y$: $c_0(y)$ is its sector, $c_L(y)$ its finest class, and $c_{\ell-1}(y)$ is the parent of $c_\ell(y)$. For the
endpoint $Y$ of a trajectory, $c_\ell(Y)$ is \emph{the endpoint's class at level $\ell$}, the class that the trajectory must eventually choose at that
level. Equality $c_\ell(y)=c_{\ell-1}(y)$ means that the class containing $y$ is not split at level $\ell$; there is then nothing to decide at that
level on this branch.

\paragraph{Classes are fixed by the target.} Classes are computed from the equilibrium measure $\mu$ alone, before any trajectory is generated
(Sec.~\ref{sec:levels}). They are therefore a property of the target, not of the generator. Changing $\beta_J$ changes $\mu$ and can change the
classes and the hierarchy (Sec.~\ref{sec:temperature}).

\paragraph{Sectors and the mirror choice.} Level~$0$ splits the measure into two mirror halves. In zero field---the only case considered here---the
measure is symmetric under spin inversion, $\mu(y)=\mu(-y)$. With $g$ one of the two ground states, the \emph{sector} of a configuration is
$s(y)=\operatorname{sign}(y\cdot g)$, and the two \emph{sectors} are the two classes of level~$0$, each of mass $\tfrac12$. A level and a sector are
therefore not the same thing: a level is a partition, and a sector is one of the two classes of level~$0$. Choosing a sector breaks the
spin-inversion symmetry; we call this choice the \emph{mirror choice}. Every ordered Ising model, a ferromagnet included, has this level, so it
carries no information about glassiness. Statements about glassy structure therefore concern the levels inside a sector, which we call
\emph{glassy levels} (levels $1$ to $L$).

\paragraph{Alignment and overlaps.} To compare configurations from the two sectors, we flip every configuration of the sector of $-g$ as a whole and
work with the \emph{aligned} configuration $\tilde y=s(y)\,y$, which always lies in the sector of $g$. Since spin inversion maps $\mu$ onto itself,
nothing is lost by doing so. This is the only gauge choice in the paper: all magnetizations, overlaps and classes inside a sector refer to aligned
configurations, and the two sectors carry mirror-image copies of the same tree. The \emph{overlap} of two configurations is $Q(y,y')=\tilde
y\cdot\tilde y'/N$; it is $1$ for identical configurations (or mirror images) and close to $0$ for unrelated ones. The \emph{magnetization} of a class
$c$ is the mean aligned configuration in that class, $\mu_c=\E_\mu[\tilde Y\mid Y\in c]$, a vector with one component per spin, and its \emph{mass}
is $\omega_c=\mu(c)$. The \emph{mean overlap within} a class $c$ is the average of $Q$ over two configurations drawn independently from $\mu$
restricted to $c$; it equals $\|\mu_c\|^2/N$. The \emph{mean overlap between} two classes $c$ and $c'$ equals $\mu_c\cdot\mu_{c'}/N$. A partition into
classes is informative when configurations in the same class overlap more than configurations in different classes.

\paragraph{Parents, children and their magnetizations.} Parents and children are \emph{classes} (including many configurations). A split takes one class
$P$, the \emph{parent}, and divides the configurations it contains into two disjoint classes, the \emph{children}, which together contain every
configuration of $P$. Like every class, the parent and the children have a mass and a magnetization. The magnetization of a class $c$ is the vector
\begin{equation}\label{eq:magnetization}
	\mu_c=\E_\mu\big[\tilde Y\mid Y\in c\big]=\frac{1}{\omega_c}\sum_{y\in c}\mu(y)\,\tilde y\ \in[-1,1]^N ,
\end{equation}
whose $i$-th component is the average aligned value of spin $i$ over the configurations of $c$, weighted by the Gibbs measure. We write $\mu_P$
for the magnetization of the parent and $\mu_1,\mu_2$ for those of its two children. Because the children divide the parent,
$\mu_P=\omega_1\mu_1+\omega_2\mu_2$, with the children's masses $\omega_{1,2}$ normalized within $P$.

\paragraph{How a split is made: cutting the hypercube.} Geometrically, each spin configuration $y\in\{\pm1\}^N$ is a corner of the hypercube
$[-1,1]^N\subset\mathbb R^N$, and the aligned configurations $\tilde y$ are the corners that lie in the sector of $g$. A magnetization
\eqref{eq:magnetization} is a weighted average of corners, so it is a point \emph{inside} the cube: the Gibbs-weighted centre of mass of the corners
that make up the class. A split of $P$ is a cut of its corners by a plane. Given the two children's centres of mass $\mu_1$ and $\mu_2$, the plane is
the one that bisects the segment between them at a right angle, and each corner of $P$ goes to the child on whose side it lies, that is, to the
nearer centre:
\begin{equation}\label{eq:hyperplane}
	y\in\text{child }1\iff \|\tilde y-\mu_1\|<\|\tilde y-\mu_2\|\iff(\mu_1-\mu_2)\cdot\tilde y>\tfrac12\big(\|\mu_1\|^2-\|\mu_2\|^2\big),
\end{equation}
and $y\in\text{child }2$ otherwise. The definition is self-consistent: $\mu_1$ and $\mu_2$ must be the magnetizations of the very children that the
plane defines. The two-means procedure of Sec.~\ref{sec:levels} finds such a pair by alternating the two steps---assign the corners to the nearer
centre, then recompute the centres---until nothing changes. Applying the rule once per split along a branch, a class of level $\ell$ is the set of
corners of its sector that lie on prescribed sides of at most $\ell$ such planes.

\paragraph{What a class looks like.} 
A useful picture, which is approximately but not exactly right, is that two sibling
classes differ by a block of spins with opposite typical values. The components of a class's magnetization with $|\mu_{c,i}|$ close to one are spins
that are nearly frozen inside the class, and components close to zero are spins that still fluctuate. For example, the magnetizations of the two
glassy classes of Fig.~\ref{fig:anatomy}a have opposite signs on $8$ of the $24$ spins. Classifying configurations by these $8$ spins alone assigns
$92\%$ of the probability mass to the correct class, but only $36\%$ of the configurations of the smaller class and $76\%$ of those of the larger one
match their class's pattern on all $8$ spins. The block summarizes the split; the plane \eqref{eq:hyperplane} defines it.

\paragraph{Split variance.} A split of a parent $P$ into two children with masses $\omega_{1,2}$ (normalized within $P$, so that
$\omega_1+\omega_2=1$) and magnetizations $\mu_{1,2}$ is characterized by its \emph{split variance}, the part of the parent's variance that the
split explains,
\begin{equation}\label{eq:splitvar}
\Delta=\omega_1\omega_2\,\|\mu_1-\mu_2\|^2\;\le\;\lambda_P ,
\end{equation}
where $\lambda_P$ is the leading eigenvalue of the parent's correlation matrix $\mathrm{Cov}_{\mu(\cdot|P)}(Y)$. The bound is the law of total variance:
the parent's correlation matrix is the sum of the average correlation matrix within the children and the between-children matrix
$\omega_1\omega_2(\mu_1-\mu_2)(\mu_1-\mu_2)^\top$, whose only nonzero eigenvalue is $\Delta$. Two examples show what $\Delta$ measures. (i) Split the
configurations according to the sign of a single spin $y_i$ that is uncorrelated with the others. The children then differ only in the $i$-th
component of their magnetization, which is $+1$ in one child and $-1$ in the other, so $\|\mu_1-\mu_2\|^2=4$ and $\Delta=4\omega_1\omega_2\le1$.
(ii) If instead the children differ by a block of $n$ spins that flip together, each such spin contributes up to $4$ to $\|\mu_1-\mu_2\|^2$, and
$\Delta$ grows in proportion to $n$ (Sec.~\ref{sec:anatomy} works out an example). A split involving a finite fraction of the
spins has $\Delta=O(N)$. We call a split \emph{collective} if $\Delta\ge2$, that is, if it accounts for at least as much variance as two independent
spins. For the sector the parent is the whole measure, and $\lambda_P=\lambda_1$ is the leading eigenvalue of the equilibrium correlation matrix
$C_0$, which is of order $N$ in an ordered system.

\subsection{Partial commitments: the state of a trajectory}\label{sec:partial}
\paragraph{Definition.} At clock $\gamma$ -- introduced in Section \ref{sec:clock} -- the generator's knowledge of its endpoint is the posterior \eqref{eq:posterior}. Summed over the
configurations of a class, it gives the \emph{partial commitment} of the trajectory to that class, and dividing by the partial commitment to the
parent gives the \emph{conditional} partial commitment:
\begin{equation}\label{eq:partial}
p_c(\gamma)=\Prb(Y\in c\mid H_\gamma),\qquad q_c(\gamma)=\frac{p_c(\gamma)}{p_{\pi(c)}(\gamma)}=\Prb\big(Y\in c\mid H_\gamma,\,Y\in\pi(c)\big).
\end{equation}
These are defined for every class $c$ of every level. $p_c$ is the probability that the trajectory currently assigns to ending in class $c$. $q_c$ is
the probability of ending in $c$ given that it ends in the parent $\pi(c)$, and so measures how far the trajectory has chosen among the children of
$\pi(c)$, whatever its state at the other levels. Along the endpoint's branch \eqref{eq:branch} we use the short notation
\begin{equation}\label{eq:branchcommit}
p_\ell(\gamma)=p_{c_\ell(Y)}(\gamma),\qquad q_\ell(\gamma)=q_{c_\ell(Y)}(\gamma)=\frac{p_\ell(\gamma)}{p_{\ell-1}(\gamma)},\qquad
p_\ell=q_0\,q_1\cdots q_\ell ,
\end{equation}
with $q_0=p_0$ the partial commitment to the endpoint's sector. Thus $p_\ell$ is the trajectory's probability for the endpoint's class at level
$\ell$, $q_\ell$ is the same probability given the endpoint's class at level $\ell-1$, and $p_\ell$ is the product of the level-by-level choices.
At $\gamma=0$ every $p_c$ equals the equilibrium mass $\omega_c$, because nothing has been observed yet; as $\gamma\to\infty$ every $p_\ell$ goes to
one and the partial commitments to all classes off the endpoint's branch go to zero.

\paragraph{Evolution.} Between these limits the partial commitments obey the equations of Gaussian filtering \citep{liptser2001statistics},
\begin{equation}\label{eq:filter}
dp_c=p_c\,(\yh_c-\yh)\cdot\big(dH_\gamma-\yh\,d\gamma\big),\qquad \yh_c=\E[Y\mid H_\gamma,Y\in c],\quad \yh=\E[Y\mid H_\gamma]=\sum_cp_c\,\yh_c ,
\end{equation}
where the last sum runs over the classes of any one level. Here $\yh$ is the posterior mean of the endpoint, the quantity that drives the generator
\eqref{eq:drift}, and $\yh_c$ is the posterior mean given that the endpoint lies in class $c$; at $\gamma=0$, $\yh_c$ is the magnetization of class $c$.
The increment $dH_\gamma-\yh\,d\gamma=d\bar B_\gamma$ is the part of the new observation that the generator could not predict (the innovation of
Proposition~\ref{prop:sl}). Three properties follow.
\begin{enumerate}[label=(\alph*),leftmargin=*]
\item \emph{Partial commitments are martingales.} The innovation has mean zero, so on average $p_c$ does not drift. Consequently they are
calibrated: among trajectories that assign probability $p$ to a class at some moment, a fraction $p$ end in that class.
\item \emph{They move only along contrasts.} The size of the move is proportional to the \emph{contrast} $\yh_c-\yh$, the difference between the
generator's estimate of the configuration within class $c$ and its overall estimate. A class whose configurations look, on average, like all
others cannot be told apart by the observations and is not decided.
\item \emph{A single trajectory fluctuates.} Because the innovation is random, an individual $p_c$ can rise and then fall back before the
trajectory finally commits.
\end{enumerate}

\paragraph{The posterior mean and its covariance.} The posterior mean $\yh$ is the order parameter of the generator, and its fluctuations are
described by the posterior covariance $C_\gamma=\mathrm{Cov}(Y\mid H_\gamma)=\partial\yh/\partial h$. Known results of stochastic localization and
of Gaussian channels \citep{eldan2013thin,elalaoui2022sampling,guo2005mutual,elalaoui2022information} state that
\begin{equation}\label{eq:known}
d\yh=C_\gamma\,d\bar B_\gamma,\qquad \frac{d}{d\gamma}\E\frac{\|\yh\|^2}{N}=\E\frac{\operatorname{Tr}C_\gamma^2}{N},\qquad
\frac{d\I(Y;H_\gamma)}{d\gamma}=\frac12\E\|Y-\yh\|^2 .
\end{equation}
In words: the posterior mean moves along the directions in which the posterior still fluctuates. The rate at which it localizes is the
spin-glass susceptibility of the tilted measure, which at $\gamma=0$ equals $N\langle Q^2\rangle$ and is set by the overlap distribution, and
information about the endpoint is gained at a rate equal to half the remaining mean-square error. We use $C_\gamma$ mainly through its leading
eigenvector, which shows which contrast is currently being decided.

\paragraph{Commitments, recrossings and events.} A trajectory \emph{commits} at level $\ell$ when $p_\ell$, its partial commitment to the
endpoint's class at that level, first exceeds $0.9$; by property (a), it then ends in a different class with probability at most $0.1$. It
\emph{recrosses} when $p_\ell$ later falls below $0.8$; the gap between the two thresholds prevents small fluctuations around $0.9$ from being
counted. A trajectory commits once at every level at which its branch splits, and never at a level where $c_\ell(Y)=c_{\ell-1}(Y)$. Since
$p_\ell\le p_{\ell-1}$, it never commits to a class before committing to that class's parent, so its commitment times are ordered along the branch.
Commitments whose values of $\gamma$ lie within a factor $f$ of each other are counted as one \emph{distinct event} ($f=2$ by default; $f=1.5$ and $3$
as checks). How a trajectory descends its tree is thus visible in its $q_\ell$: levels whose $q_\ell$ rise in separate intervals of $\gamma$ are
decided one after another, and levels whose $q_\ell$ rise together are decided as one compound decision (Fig.~\ref{fig:partial}a--c).

\subsection{The ensemble view: information per level}\label{sec:levelrate}
Averaging over trajectories turns partial commitments into information. Given the endpoint's class at level $\ell-1$, the uncertainty about its
class at level $\ell$ is the entropy $\Hh(C_\ell\mid C_{\ell-1})$ of the equilibrium masses of the children. Along a trajectory this uncertainty
becomes the entropy of the conditional partial commitments $q_c$, which shrinks as the trajectory decides. The \emph{information about level
$\ell$}, beyond what is known about level $\ell-1$, is the average decrease,
\begin{equation}\label{eq:levelinfo}
\I(C_\ell;H_\gamma\mid C_{\ell-1})=\Hh(C_\ell\mid C_{\ell-1})-\E\,\Hh(C_\ell\mid C_{\ell-1},H_\gamma),
\end{equation}
which rises from $0$ at $\gamma=0$ to $\Hh(C_\ell\mid C_{\ell-1})$ as $\gamma\to\infty$. The following proposition gives an explicit expression for the mutual information (per level) increase.  

\begin{proposition}[Level-resolved information rate]\label{prop:levelrate}
For nested classes fixed before the trajectories are generated,
\begin{equation}\label{eq:levelrate}
\frac{d}{d\gamma}\I(C_\ell;H_\gamma\mid C_{\ell-1})=\frac12\,\E\sum_{c\in C_\ell}p_c\,\big\|\yh_c-\yh_{\pi(c)}\big\|^2\;\ge0 ,
\end{equation}
and along every trajectory the conditional class entropy is a supermartingale whose compensator is the same contrast term. Summed over levels,
$\tfrac N2(1-\bar q)=\sum_\ell R_\ell(\gamma)+R_{\rm within}(\gamma)$, with $R_\ell$ the right-hand side of \eqref{eq:levelrate},
$\bar q=\E\|\yh\|^2/N$ and $R_{\rm within}=\tfrac12\E\operatorname{Tr}\mathrm{Cov}(Y\mid H_\gamma,C_L)$.
\end{proposition}
Notice that the statement is equivalent to the conditional form of the I--MMSE relation \citep{duncan1970calculation,guo2005mutual}. It is the ensemble counterpart of property (b):
information about level $\ell$ is gained exactly at the rate at which the contrasts between the classes of that level and their parents are
exposed. The last statement of the proposition is a \emph{budget}: the total information rate of the generator, $\tfrac N2(1-\bar q)$, is shared among the levels and the
local detail inside the finest classes, and the proposition says exactly how (Appendix~\ref{app:tools}).

\paragraph{Decisions, commitments and decision times.} We use one word, \emph{decide}, for one idea throughout the paper: a split of a parent
class into two children is decided when the generator's state no longer leaves open which of the two children contains its endpoint. The classes
against which this is measured are always those fixed by the target (Sec.~\ref{sec:defs}); what evolves on the clock $\gamma$ is the state of the
generator along its trajectory. The idea is quantified at two levels of description. For a \emph{single trajectory}, the split is decided at its
\emph{commitment}, defined above.
For the \emph{ensemble} of trajectories, it is decided at its \emph{decision time} $\gamma_{1/2}$, the value of $\gamma$ at which the information
about the level reaches half of its final value $\Hh(C_\ell\mid C_{\ell-1})$; the \emph{window} of the split is the range of $\gamma$ over which the
information rises from $10\%$ to $90\%$ of that value. Since the information is the average over trajectories of the decrease of the entropy of
their conditional partial commitments, individual trajectories commit during the window, mostly in its later part, because a commitment requires a
probability of $0.9$ rather than one half. Because the final value is computed exactly, decision times do not depend on how long the trajectories are
observed. Individual spins are decided in the same sense---the posterior magnetization $\yh_i$ of spin $i$ approaches $\pm1$---typically at
$\gamma\gtrsim1$, a part of the path that we call the \emph{local stage}.

\subsection{When is a split decided? The binary-channel prediction}\label{sec:channel}

\paragraph{What is being predicted, and the status of the prediction.} This subsection asks when, on the clock $\gamma$, the generator decides a
given split of the target's classes, and answers with a \emph{prediction} of the decision time computed from the target alone, before any
trajectory is generated. Its status should be clear from the outset. It is not an assumption of the analysis: the classes, the trajectories, the
partial commitments and the measured decision times are all computed without it. It is an approximate formula, obtained from an exact reduction
(the next paragraph) and one idealization of each split (the paragraph after it). How accurate the idealization is, and why it fails where it
fails, are measured results of the paper (Secs.~\ref{sec:lawexp} and \ref{sec:families}).

\paragraph{Reduction to one parent.} At clock $\gamma$ the generator knows as much about its endpoint as an observer who measures $Y$ through additive
Gaussian noise, $H_\gamma=\gamma Y+B_\gamma$ -- which is a \emph{Gaussian channel}. Knowing the endpoint's class at level $\ell-1$ only changes the prior of
this observer, from $\mu$ to the parent's own Gibbs measure $\mu(\cdot\mid P)$. Hence, averaging over the parents $P\in C_{\ell-1}$ with their
masses,
\begin{equation}\label{eq:condprior}
\I(C_\ell;H_\gamma\mid C_{\ell-1})=\sum_{P\in C_{\ell-1}}\omega_P\,\I_{\mu(\cdot\mid P)}(C_\ell;H_\gamma),
\end{equation}
where $\I_{\mu(\cdot|P)}$ is the information of the same channel when the prior is $\mu(\cdot\mid P)$. Below each parent, the conditional partial
commitments $q_c$ therefore evolve exactly like the partial commitments of a generator whose target is the parent's own Gibbs measure. Every split,
at every depth, can thus be analysed as if it were the first split of a smaller problem.

\paragraph{A binary-channel benchmark.}
Eq.~\eqref{eq:condprior} is exact; to obtain a simple, testable prediction we now \emph{idealize} each child's conditional distribution by a point mass at its magnetization and initially assign the two children equal weights. If the resulting child contrast dominates the parent covariance and is aligned with its leading eigenvector $v_P$, the two projected means are separated by approximately
$2\sqrt{\lambda_P}$. Thus the scalar observation
$\sqrt\gamma\,v_P\cdot Y+\xi$, with $\xi\sim\mathcal N(0,1)$, reduces to a balanced binary Gaussian channel with parameter $s=\gamma\lambda_P$. Its information is
$I_{\rm bin}(s)=\log2-\E\log\big(1+e^{-2(s+\sqrt{s}\,\xi)}\big)$,
which reaches $10\%$, $50\%$ and $90\%$ of $\log2$ at
$s=0.149,1.044,3.76$. Hence the benchmark predicts
\begin{equation}\label{eq:law}
\gamma_{1/2}^{(P)}\simeq\frac{1.044}{\lambda_P},
\end{equation}
with a $10$--$90\%$ window spanning a factor of approximately $25$
in $\gamma$.

The channel calculation is exact for two equally weighted point masses,
not for the full conditional Gibbs distributions of finite cells.
Although the constructed splits are binary by definition, their children
can have substantial internal fluctuations or unequal masses, and the
leading covariance direction may not represent their contrast.
Eq.~\eqref{eq:law} is therefore a \emph{benchmark}, compared with
the exact information curves in Secs.~\ref{sec:lawexp} and
\ref{sec:families}; departures quantify where the approximation fails.
The inverse-leading-eigenvalue scale was identified by
\citet{biroli2024dynamical}; the balanced channel gives its numerical
constant and decision window and, through~\eqref{eq:condprior}, applies
to each parent. Related work derives successive speciation scales from
free-entropy differences~\citep{achilli2026theory}. Replacing
$\lambda_P$ by successive eigenvalues of $C_0$ gives a label-free
estimate when the level contrasts align with those eigenvectors.

\paragraph{Unequal child masses.}
Eq.~\eqref{eq:law} assumes that the two effectively rigid children have equal conditional masses, in addition to being approximately aligned with the leading covariance mode. For child masses $w,1-w$ and centroid separation $d=\|\mu_1-\mu_2\|$, put $g=\gamma d^2$ and $a=\log[w/(1-w)]$. The log posterior odds $L$ in the rigid two-center channel satisfy
\begin{equation}\label{eq:unbalanced_binary}
 L\mid C=1\sim{\cal N}(a+g/2,g),\qquad L\mid C=2\sim{\cal N}(a-g/2,g),
\end{equation}
so $I_{\rm bin}(g,w)=H_b(w)-\mathbb E[H_b((1+e^{-L})^{-1})]$, with $H_b$ the binary entropy and expectation over the mixture in~\eqref{eq:unbalanced_binary}. Its half-information clock solves $I_{\rm bin}(g_{1/2},w)=H_b(w)/2$; for $w=1/2$, $g_{1/2}=4(1.044)$ and the balanced rule is recovered because $\lambda_P\simeq d^2/4$. At small $w$, replacing this channel by Eq.~\eqref{eq:law} can strongly overestimate its decision time. Appendix~\ref{app:additional_validation} tests this refinement on the original exact instances.

\paragraph{Larkin balance and decision scales.}
The original Larkin argument \citep{Larkin1970Inhomogeneities,LarkinOvchinnikov1979Pinning} balances disorder-induced distortions against elastic restoring
forces. Garnier-Brun \emph{et al.}~\citep{GarnierBrun2026Foam} adapt this idea to competing low-lying states in a high-dimensional random landscape, comparing their
characteristic free-energy gap $E_{\rm gap}(\ell)$ at separation $\ell$ with Gaussian pinning: $\gamma\ell^2\sim E_{\rm gap}(\ell)$. Their model-specific gap statistics then predict a cascade of progressively resolved scales.

For our finite spin glasses, an analogous competition can be examined at each parent--children split. For approximately rigid children with relative masses
$w$ and $1-w$ and centroid separation $d$, the Gaussian log-odds channel in Eq.~\eqref{eq:unbalanced_binary} is governed by $g=\gamma d^2$ and the prior free-energy bias $a=\log[w/(1-w)]$. Balanced children become distinguishable at $g=O(1)$, while unequal weights shift the decision scale. Unlike the model of~\citet{GarnierBrun2026Foam}, no specific gap--distance scaling is assumed here in our study. The resulting decision scales are expressed in the universal localization clock $\gamma$, independently of the isotropic PID protocol. 
 Unlike the binary-channel benchmark, the Larkin balance itself does not predict a complete information curve. Therefore, and without an independently established gap--distance law for our finite cells, we view the Larkin balance as a useful physical interpretation of the resolution scale.

\paragraph{Four consequences.} The prediction has four consequences that organize the experiments.
\begin{enumerate}[label=(\arabic*),leftmargin=*]
\item \emph{Collectivity.} Individual spins are decided at $\gamma\sim1$. A split is decided before them only if its $\lambda_P$, and hence its
split variance $\Delta\le\lambda_P$, is larger than about $2$; this is why we require $\Delta\ge2$.

\item \emph{Order.} For approximately balanced, rigid and spectrally aligned splits, the parent leading eigenvalue determines the half-information time and a factor around $25$ in eigenvalue separates non-overlapping $10$--$90\%$ windows. Depth alone does not order these decisions. With unequal masses the ordering additionally depends on the child weights and contrasts, as quantified in Eq.~\eqref{eq:unbalanced_binary}. In an ordered glass the mirror sector normally comes first; close to the transition, its spectral direction may fail the rigid-binary approximation and the sector decision can be delayed.

\item \emph{Hand-over.} Once a parent is decided, the posterior covariance $C_\gamma$ approaches the correlation matrix of that parent, so its leading
eigenvector---equivalently, that of the Jacobian $\partial\yh/\partial h$ of a learned denoiser---turns from the parent's contrast to the contrast
between its children.

\item \emph{Timing alone is not evidence of a hierarchy.} The spectral benchmark alone does not establish a hierarchy. A correctly predicted or well separated window
does not by itself show that there are nested classes: a mixture of two states has a sector decision and, much later, an arbitrary split of its
thermal noise, in widely separated windows. Classes and partial commitments are needed in addition.
\end{enumerate}

\paragraph{Crossovers, not phase transitions.} As a function of $\log\gamma$, the curve $I_{\rm bin}(\gamma\lambda_P)$ has a fixed shape. A larger $N$
increases $\lambda_P$ and shifts the window to smaller $\gamma$, but does not make it narrower. Deciding a split is therefore a crossover, not a
phase transition along the generator's path. If $\lambda_P$ grows in proportion to $N$, all macroscopic splits are decided at $\gamma\to0^+$ in the
limit $N\to\infty$; this is the familiar statement that an infinitesimal field selects a state. A decision at finite $N$ is the same selection,
spread out in time on the scale $\tau=N\gamma$. Genuine phase transitions along the path \citep{ghio2024sampling} require exponentially many
competing alternatives and would appear as non-analytic behaviour of $\bar q(\gamma)$ at $\gamma=O(1)$; we see none, and could not detect one at
$N\le24$.

\FloatBarrier
\section{Identifying the classes}\label{sec:levels}
Decisions are measured against classes that are fixed before any trajectory is generated, so that the classes are not read off the dynamics they
are meant to explain. For $N\le24$ the classes are computed from the \emph{exact} Gibbs measure, without sampling. They are also computed without
using the eigenvectors of the correlation matrices, so that the binary-channel prediction, which uses the leading eigenvalues, is tested rather than
built in. The construction has five steps.
\begin{enumerate}[leftmargin=*]
\item \emph{Sectors.} Level~$0$ consists of the two sectors, $s(y)=\operatorname{sign}(y\cdot g)=\pm1$, with $g$ the ground state (ties broken by the
sign of $y_1$). All further splits are computed for the aligned configurations $\tilde y=s(y)\,y$ (Sec.~\ref{sec:defs}), so the two sectors carry the
same tree. Because $J=\beta_JJ^{(0)}$, the ground state, and hence the alignment, does not depend on $\beta_J$.
\item \emph{Candidate splits.} Each class is split by an exact, mass-weighted two-means procedure. Starting from two trial magnetizations, every
configuration of the class is assigned to the nearer one by the hyperplane rule \eqref{eq:hyperplane}; the two magnetizations are then recomputed
exactly as the Gibbs averages over the two resulting children, and the two steps are repeated until the partition stops changing. The procedure is
started from several initial pairs---the two most probable zero-temperature valleys in the class, random pairs of configurations drawn from the
Gibbs measure, and a pair displaced along the class's principal direction---and the split with the largest split variance is kept. By
\eqref{eq:levelrate}, this is the split about which the generator initially gains information fastest: at $\gamma=0$ the rate is
$\tfrac12\sum_c\omega_c\|\mu_c-\mu_{\pi(c)}\|^2$.
\item \emph{Acceptance.} A candidate split is accepted only if it passes three tests: both children carry at least $3\%$ of the measure; the mean
overlap within the children exceeds the mean overlap between them by at least $0.10$; and the split is collective, $\Delta\ge2$. The last test is
essential. Without it, the split according to a single spin passes the overlap test at $N\le20$ (its overlap gap is $2/N$), and even a paramagnet at
$N=16$ would yield a tree three levels deep.
\item \emph{Depth.} Splitting stops when no class has an accepted split, or after three levels below the sector. A class without an accepted split
reappears unchanged at the next level (Sec.~\ref{sec:defs}).
\item \emph{Annotation.} Each accepted split is labelled by the zero-temperature valleys on its two sides. A zero-temperature valley is the set of
configurations that reach the same local energy minimum by single spin flips that lower the energy. A split that separates different valleys is a
split between metastable states; a split within one valley would be a split of thermal fluctuations, which we would not call glassy.
\end{enumerate}
For $N=32$ and $64$, where the measure cannot be enumerated, the sector and the first glassy split are identified from equilibrium samples by
recursive bisection along the leading principal component, and they are accepted only if the same partition is found on an independent set of
samples, as measured by the adjusted Rand index (Sec.~\ref{sec:procedure}).

\paragraph{An independent test of the operational cells.}
The mass, mean-separation and collectivity gates identify finite-size classes relevant for collective generative decisions. However, in general these objects only provide a reduced description of the terminal hierarchy.  In particular, for independent replicas $Y,Y'$ conditioned on classes $A_\alpha,A_\beta$, the overlap $q_N=N^{-1}Y\cdot Y'$ is an additionl useful object that can retain nonvanishing conditional variance. Following a suggestion of M.~M\'ezard (private communication), we assess how much of the genuine overlap distribution is lost when a cell is represented only by its magnetization $m_\alpha=\E(Y\mid A_\alpha)$. With $w_\alpha=\mu(A_\alpha)$ and $q_{\alpha\beta}=N^{-1}m_\alpha\cdot m_\beta$, define
\begin{equation}\label{eq:overlap_residual}
 R_{J,N}:=\E(q_N^2)-\sum_{\alpha,\beta}w_\alpha w_\beta q_{\alpha\beta}^2
 =\sum_{\alpha,\beta}w_\alpha w_\beta\operatorname{Var}(q_N\mid A_\alpha,A_\beta)\ge0.
\end{equation}
Note, that the residual is entirely a static property of the target and the chosen partition: it involves neither the localization clock nor the PID dynamics. It tests whether the operational classes preserve equilibrium overlap statistics, independently of how and when the generator commits to them. 

For the complete spin-inversion-symmetric partition, set $C_0=\E(YY^\top)$, $M=\sum_\alpha w_\alpha m_\alpha m_\alpha^\top$ and $W=C_0-M$. Independence of the replicas and the covariance decomposition give the exact identity
\begin{equation}\label{eq:overlap_matrix}
 R_{J,N}=\frac{\operatorname{Tr}(C_0^2)-\operatorname{Tr}(M^2)}{N^2}
 =\frac{2\operatorname{Tr}(MW)+\operatorname{Tr}(W^2)}{N^2}.
\end{equation}
In particular, refinements never increase $R_{J,N}$, but singleton cells make it zero trivially. Vanishing residual overlap variance (under appropriate thermodynamic-limit and disorder-averaging assumptions) is thus a necessary overlap-concentration check, not a sufficient characterization of pure states. Appendix~\ref{app:additional_validation} reports the corresponding finite-size diagnostic.

\section{Experiments}\label{sec:experiments}
The experiments answer a sequence of questions, each prompted by the answer to the previous one. We first ask whether, once the generator has
chosen a sector, the measure contains further collective classes, and how the generator decides among them (Sec.~\ref{sec:anatomy}). We then ask
whether a single trajectory makes one decision, one decision per level, or something in between (Sec.~\ref{sec:paths}). Next we ask whether the
classes behind these decisions are a robust property of the glass, by varying the temperature (Sec.~\ref{sec:temperature}) and the system size
(Sec.~\ref{sec:scan}). We then test whether the decision times can be predicted from equilibrium spectra alone (Sec.~\ref{sec:lawexp}) and compare
sparse and dense glasses (Sec.~\ref{sec:families}). Finally we ask what changes when the generator takes finitely many steps
(Sec.~\ref{sec:protocol}), whether the procedure could have produced the same picture for targets without glassy structure (Sec.~\ref{sec:controls}),
and whether the picture survives beyond exact enumeration and in trained denoisers (Sec.~\ref{sec:larger}). Each subsection follows the same
pattern: it states the question and the data used, goes through the corresponding figure panel by panel, and ends with the conclusion we draw.

\subsection{Procedure, systems and controls}\label{sec:procedure}
\paragraph{How each instance is analysed.} Every instance is analysed in the same six steps, summarized in Table~\ref{tab:procedure}.
\begin{enumerate}[leftmargin=*]
\item \emph{Exact measure.} For $N\le24$ the Gibbs measure \eqref{eq:target} is computed exactly, by enumerating all $2^N$ configurations
(about $1.7\times10^7$ at $N=24$).
\item \emph{Classes.} The classes are identified from the exact measure by the procedure of Sec.~\ref{sec:levels}, before any trajectory is generated.
\item \emph{Predicted decision times.} For every split, the binary-channel prediction \eqref{eq:law} gives the decision time from the leading eigenvalue
$\lambda_P$ of the parent's correlation matrix. This prediction uses only the equilibrium measure.
\item \emph{Trajectories.} Trajectories are generated \emph{endpoint first} (in the stochastic-localization literature this is called
\emph{planting}). An endpoint $Y$ is drawn from the target $\mu$ and fixed; an independent
Brownian motion $B$ is drawn; and the field process is set to $H_\gamma=\gamma Y+B_\gamma$. By Proposition~\ref{prop:sl}, the pair $(Y,H)$ obtained in
this way has exactly the same joint law as the endpoint of the PID generator started at $x=0$ and its field process, for every protocol. Equivalently,
$X_t=H_{\gamma(t)}/\kappa(t)$ is a path that starts at $X_0=0$ and ends at $X_1=Y$, distributed as the Gaussian bridge \eqref{eq:bridge}. Nothing is
imposed by hand: the trajectories are fully determined by $\mu$ and are genuine trajectories of the generator, merely generated in the order
``endpoint, then path''. One trajectory serves all protocols at once, because the protocol only relabels $\gamma$ by $t$. Each endpoint receives
one trajectory; the number of trajectories per instance is our choice of sample size (see below).
\item \emph{Partial commitments along each trajectory.} At every value of $\gamma$ on a logarithmic grid, the exact posterior \eqref{eq:posterior} gives
all partial commitments $p_c(\gamma)$, from which we obtain commitments, distinct events and recrossings (Sec.~\ref{sec:partial}) and, by averaging
over trajectories, the information about each level and its measured decision time (Sec.~\ref{sec:levelrate}).
\item \emph{Comparison.} Measured and predicted decision times are compared. Both refer to the same classes, which were fixed in step~2.
\end{enumerate}

\paragraph{How the endpoints are chosen.} Endpoints are drawn from the target $\mu$, but in a \emph{stratified} way, so that rare classes are not
missed. Before sampling, each finest class $k$ (mass $\omega_k$) is allotted $n_k=\max\big(2,\operatorname{round}(n\omega_k)\big)$ endpoints, and these are
drawn from $\mu$ restricted to that class. Each endpoint in class $k$ carries the weight $\omega_k/n_k$, so that weighted averages over the trajectories
are unbiased averages over $\mu$. We use $n=256$, $96$ and $48$ at $N=16$, $20$ and $24$ ($128$ for the $N=24$ instance of Fig.~\ref{fig:anatomy}); because
every class receives at least two endpoints, the actual number of trajectories per instance is slightly larger than $n$. For the per-trajectory counts
of Secs.~\ref{sec:paths} and \ref{sec:families} we use $128$ trajectories per instance at $N=16$ and $64$ at $N=20$ (Appendix~\ref{app:methods}).

\begin{table}[!htbp]\centering\small
\begin{tabular}{@{}p{0.27\linewidth}p{0.4\linewidth}p{0.27\linewidth}@{}}\toprule
Quantity & How it is computed & Theory used \\ \midrule
classes & exact two-means splits of $\mu$, gates on mass, overlap gap, $\Delta\ge2$ & Eq.~\eqref{eq:splitvar}, Prop.~\ref{prop:levelrate} \\
predicted decision time & half point of $I_{\rm bin}(\gamma\lambda_P)$, averaged over parents & Eqs.~\eqref{eq:condprior}, \eqref{eq:law} \\
partial commitments & exact posterior along $H_\gamma$ & Prop.~\ref{prop:sl}, Eqs.~\eqref{eq:posterior}, \eqref{eq:partial} \\
commitments, events, recrossings & thresholds on $p_\ell=p_{c_\ell(Y)}$ along one trajectory & Eq.~\eqref{eq:filter} \\
information, decision time & $\Hh(C_\ell)-\E\,\Hh(C_\ell|H_\gamma)$, increments & Prop.~\ref{prop:levelrate} \\
hand-over of contrasts & leading eigenvector of $C_\gamma$ & Eq.~\eqref{eq:known} \\
effect of the protocol & $K$-step Euler integration, exact drift & Props.~\ref{prop:sl}, \ref{prop:energy} \\ \bottomrule
\end{tabular}
\caption{What is computed, how, and from which result of Secs.~\ref{sec:pid}--\ref{sec:decisions}.}\label{tab:procedure}
\end{table}

\paragraph{Systems.} Both glass families have independent Gaussian couplings and are studied from $\beta_J=0.5$ (paramagnet) to $\beta_J=3$ (deep in
the glass phase). \emph{Bethe glasses} \citep{viana1985phase,mezard2001bethe} have couplings $J^{(0)}_{ij}\sim\mathcal N(0,1)$ on the edges of a random
three-regular graph: every spin interacts with three others, and the graph looks locally like a tree. Their glass transition is at $\beta_c\approx1.34$,
where $2\,\E\tanh^2(\beta_cJ^{(0)})=1$. \emph{SK glasses} \citep{sherrington1975solvable} have couplings $J^{(0)}_{ij}\sim\mathcal N(0,1/N)$ between all
pairs of spins, and their transition is at $\beta_c=1$. The Bethe family is thus sparse and the SK family is dense. In the thermodynamic limit both have
a glass phase with broken replica symmetry \citep{mezard1987sgtb,mezard2001bethe}. When the two families are compared, temperature is measured by
$\beta_J/\beta_c$. All instances are \emph{unselected}: we use consecutive disorder seeds and keep every instance. There are $60$, $24$ and $12$ Bethe
instances at $N=16$, $20$ and $24$, and $30$ and $24$ SK instances at $N=16$ and $20$.

\paragraph{Larger systems.} At $N=32$ and $64$ the measure cannot be enumerated, and two ingredients of the procedure must be replaced by Monte
Carlo methods. First, the classes are identified from \emph{equilibrium samples}, that is, configurations drawn from $\mu$ by a Markov-chain Monte
Carlo method (parallel tempering); the sector and the first glassy split are found by recursive bisection along the leading principal component of
these samples, and are accepted only if they are reproduced on an independent set of samples. Second, the posterior along each trajectory is
estimated by sequential Monte Carlo instead of being computed exactly. We analyse three Bethe instances per size; unlike the exact instances, they
were \emph{selected}, for having classes that are reproducible across independent sample sets (Appendix~\ref{app:methods}).

\paragraph{Negative and positive controls.} To check that the procedure does not invent glassy structure, we apply it, unchanged, to systems in
which we know what structure to expect.
\begin{itemize}[leftmargin=*]
\item \emph{Ferromagnet on the same graphs.} We replace every coupling $J_{ij}$ of a Bethe instance by its magnitude $|J_{ij}|$, keeping the graph and
the temperature. The resulting ferromagnet has no frustration. It has a phase transition at $\beta_J\approx0.83$ (where $2\,\E\tanh(\beta_J|J^{(0)}|)=1$),
at which the spin-inversion symmetry breaks; below it the measure consists of two mirror halves, mostly up and mostly down, and no further structure is
expected at lower temperature. This control tests whether the glassy levels could be a consequence of the graph and the coupling magnitudes alone.
\item \emph{Mattis model.} $\mu(y)\propto\exp\big(\tfrac{\beta_J}{2N}(\xi\cdot y)^2\big)$, with a pattern $\xi\in\{\pm1\}^N$ drawn at random for each
instance and then fixed. The change of variables $y_i\to\xi_iy_i$ maps it onto the fully connected (Curie--Weiss) ferromagnet, so the choice of pattern
is immaterial: the model has a transition at $\beta_J=1$, and at the value $\beta_J=1.37$ used here its measure consists of two mirror states, $y$
close to $\xi$ or to $-\xi$ (magnetization along the pattern about $0.8$), with no further structure. This control tests whether a strong sector alone
can produce glassy levels.
\item \emph{Independent spins.} The uniform measure on $\{\pm1\}^N$, i.e.\ couplings $J=0$ or infinite temperature. Its correlation matrix is the
identity, so every split has $\Delta\le\lambda_P=1$, below the collectivity gate $\Delta\ge2$ of Sec.~\ref{sec:levels}, and no split can be accepted.
This control is used at $N=32,64$, where it tests the structural criteria used there, together with a \emph{two-state mixture} of two product measures with site
magnetizations $\pm0.8$, the analogue of the Mattis model.
\item \emph{Positive controls.} A mixture of $36$ product states arranged on a prescribed three-level tree (Appendix~\ref{app:E0}) checks the identities
and the pipeline; an unfrustrated mixture of four product states arranged as two mirror parents with two children each checks that a genuine nested
structure is detected when it is present.
\end{itemize}

\subsection{One instance in detail}\label{sec:anatomy}
\begin{figure}[!htbp]\centering
\includegraphics[width=\linewidth]{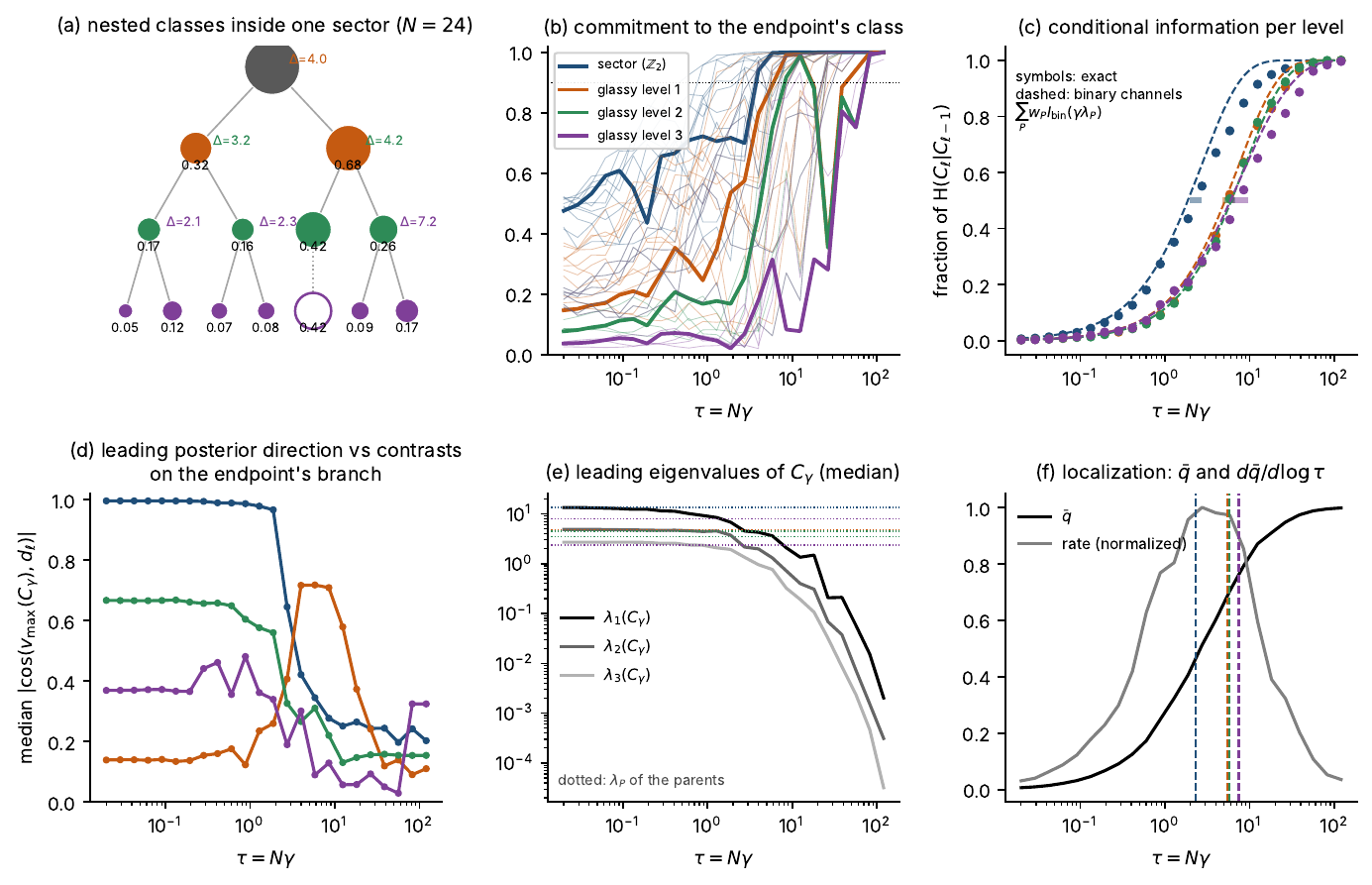}
\caption{\textbf{One $N=24$ Bethe glass at $\beta_J=2.5$}, exact enumeration, $128$ trajectories with endpoints drawn from $\mu$ (the number is our choice of sample size); time on the scaled
clock $\tau=N\gamma$.
(a) The classes inside one sector, computed from the equilibrium measure: circles are classes (numbers: masses within the sector; open circle: a
class not split further), and each split is labelled by its split variance $\Delta$.
(b) Partial commitment $p_\ell=p_{c_\ell(Y)}$ of each trajectory to the endpoint's class at each level $\ell$; thin lines: all trajectories; thick lines: one
trajectory; dotted: the commitment threshold $0.9$. (c) Information about each level, averaged over trajectories and normalized by its final value
(symbols), with the binary-channel prediction from the parents' $\lambda_P$ (dashed); bars: $68\%$ bootstrap intervals of the measured decision times.
(d) Median $|\cos|$ between the leading eigenvector of the posterior covariance $C_\gamma$ and the contrast of each level. (e) Median of the three
leading eigenvalues of $C_\gamma$; dotted: the parents' $\lambda_P$. (f) Mean self-overlap $\bar q$ and its normalized logarithmic rate; dashed: the
decision times of panel c.}\label{fig:anatomy}
\end{figure}

\emph{Question.} The generator first chooses a sector, that is, it breaks the spin-inversion symmetry (the \emph{mirror choice}, Sec.~\ref{sec:defs}).
Does the measure inside the chosen sector contain further collective classes? And if so, when and how does the generator decide among them, i.e.\ at
which values of the clock $\gamma$ do its partial commitments to these classes go from their equilibrium values to zero or one?
\emph{Data.} One $N=24$ Bethe glass at $\beta_J=2.5$ (about $1.9\beta_c$, well inside the glass phase), with $128$ trajectories.
Fig.~\ref{fig:anatomy} summarizes this instance in six panels, which we discuss in turn. Time is shown on the scaled clock $\tau=N\gamma$.

\emph{The classes (panel a of Fig.~\ref{fig:anatomy}).} The panel shows the classes found inside one sector. These classes are a property of the target: they are computed from the
equilibrium measure $\mu$---which is also the distribution of the generator's endpoints, reached as $\tau\to\infty$---by the procedure of
Sec.~\ref{sec:levels}, without using any trajectory. Each circle is a class, labelled by its mass within the sector, and each split is labelled by its
split variance $\Delta$. The sector splits into two glassy classes of masses $\omega_1=0.32$ and $\omega_2=0.68$. Their magnetizations $\mu_1$ and
$\mu_2$ have $\|\mu_1\|^2/N=0.71$ and $\|\mu_2\|^2/N=0.62$ (a typical spin has $|\mu_{c,i}|\approx0.75$ in both classes); they have opposite signs on the block of $8$ spins
discussed in Sec.~\ref{sec:defs}, so that $\|\mu_1-\mu_2\|^2/N=0.77$ and $\Delta=\omega_1\omega_2\|\mu_1-\mu_2\|^2=4.0$.
Accordingly, configurations in the same class have mean overlap $0.65$ (the mass-weighted average of $\|\mu_c\|^2/N$), and configurations in
different classes $\mu_1\cdot\mu_2/N=0.28$ (Sec.~\ref{sec:defs}). Each of the two classes
splits again ($\Delta=3.2$ and $4.2$), and three of the four resulting classes split once more ($\Delta=2.1$ to $7.2$), giving three glassy levels and
seven finest classes per sector, $14$ in total. The fourth class (mass $0.42$) has no split that passes the tests of Sec.~\ref{sec:levels}; it
therefore reappears unchanged at the finest level and is drawn as an open circle. Every split separates distinct zero-temperature valleys---sets of configurations that descend, by
single spin flips that lower the energy, to different local minima---so the classes are not thermal domains of a single valley. The leading eigenvalues
of the parents' correlation matrices, which set the predicted decision times, are $\lambda_P=13.5$ for the sector, $4.9$ for the first glassy level, and
$3.4$--$4.5$ and $2.3$--$7.9$ for the parents of the second and third glassy levels. \emph{Inside the chosen sector there is thus a three-level
hierarchy of collective classes.}

\emph{Single trajectories (panel b of Fig.~\ref{fig:anatomy}).} The panel shows, for each level $\ell$, the partial commitment
$p_\ell=p_{c_\ell(Y)}$ of every trajectory to the endpoint's class at that level \eqref{eq:branchcommit}, i.e.\ to the class that contains its
endpoint (thin lines), with one trajectory highlighted (thick lines). All partial commitments start at the equilibrium masses---$0.5$ for the
sector---and end at one. Every trajectory commits (reaches $p_\ell\ge0.9$) at every level at which its branch splits; afterwards, $9\%$ of the trajectories recross at
the sector (fall back below $0.8$ before committing again) and $4$--$6\%$ at a glassy level. The highlighted trajectory commits at every level and
recrosses once, at a glassy level, before committing for good. \emph{Individual trajectories fluctuate, but all of them commit, and
the commitments come in the order of the levels.}

\emph{The ensemble (panel c of Fig.~\ref{fig:anatomy}).} The symbols show, for each level, the information that the generator has acquired about that level, averaged over
trajectories and normalized by its final value; the dashed lines are the binary-channel predictions computed from the parents' $\lambda_P$, and the
horizontal bars are $68\%$ bootstrap intervals of the measured decision times. The sector is decided at $\gamma_{1/2}=0.096$ ($68\%$ interval
$0.089$--$0.104$), against a prediction of $0.078$. The three glassy levels are decided at $\gamma_{1/2}=0.23$, $0.24$ and $0.31$, against predictions of
$0.21$, $0.25$ and $0.26$. \emph{The binary-channel prediction is within $25\%$ of every measured decision time of this instance; the three glassy levels are decided almost
together, a factor $2.4$ after the sector, in windows that overlap with that of the sector.}

\emph{Which contrast is being decided (panels d and e of Fig.~\ref{fig:anatomy}).} Panel d shows the median, over trajectories, of $|\cos|$ between the leading eigenvector of the
posterior covariance $C_\gamma$ and the contrast $d_\ell$ of each level along the endpoint's branch, the difference $\mu_1-\mu_2$ between the magnetizations of the two children
of $c_{\ell-1}(Y)$, the endpoint's class at the level above. Up to $\tau\approx2$ the leading eigenvector is the sector contrast ($|\cos|\approx1$). It then turns away, and the first glassy
contrast takes over ($|\cos|\approx0.72$ at $\tau\approx5$--$8$), before the posterior concentrates on single configurations. The curves behave
differently because the levels are decided at different times. The sector curve can only decrease: the sector contrast is the leading direction
from the start and stops being so once the sector is decided. The first-glassy curve (orange) rises and then falls. While the sector is undecided,
the sector contrast dominates the fluctuations of the posterior; once it is decided, the first glassy contrast is the largest remaining
fluctuation, so the curve rises to its maximum at $\tau\approx6$--$10$; once the first glassy level is decided as well, its contrast loses its weight
too, and the curve falls. The second-glassy curve starts high ($0.67$) only because its contrast has a sizeable component along the sector
direction, which dominates at small $\tau$; it falls together with the sector curve and never becomes the leading direction, because the first
two glassy levels are decided together and the first glassy contrast carries more variance. (The third glassy level splits on only some branches,
and its median is not shown.) Panel e shows the three
leading eigenvalues of $C_\gamma$ (medians) together with the parents' $\lambda_P$ (dotted). At small $\tau$ they equal the eigenvalues of the
equilibrium correlation matrix; they then collapse one after another, the largest (the sector mode) first, as the corresponding splits are decided.
This is the hand-over predicted in Sec.~\ref{sec:channel}: \emph{once a parent is decided, the dominant fluctuation of the posterior is the contrast
of its children.} It is also what the Jacobian $\partial\yh/\partial h=C_\gamma$ of a trained denoiser would show (Sec.~\ref{sec:learned}).

\emph{A global measure does not separate the decisions (panel f of Fig.~\ref{fig:anatomy}).} The panel shows the mean self-overlap $\bar q=\E\|\yh\|^2/N$, which measures how
far the posterior mean has localized, and its logarithmic rate $d\bar q/d\log\tau$; the dashed lines mark the decision times of panel c. The rate has a
single broad peak around $\tau\approx2$--$5$ that covers the sector and the glassy decisions alike. \emph{The level-resolved quantities of panels b--e are
needed to see the individual decisions; the overall localization of the posterior does not reveal them.}

\subsection{How a trajectory descends the tree}\label{sec:paths}
\begin{figure}[!htbp]\centering
\includegraphics[width=\linewidth]{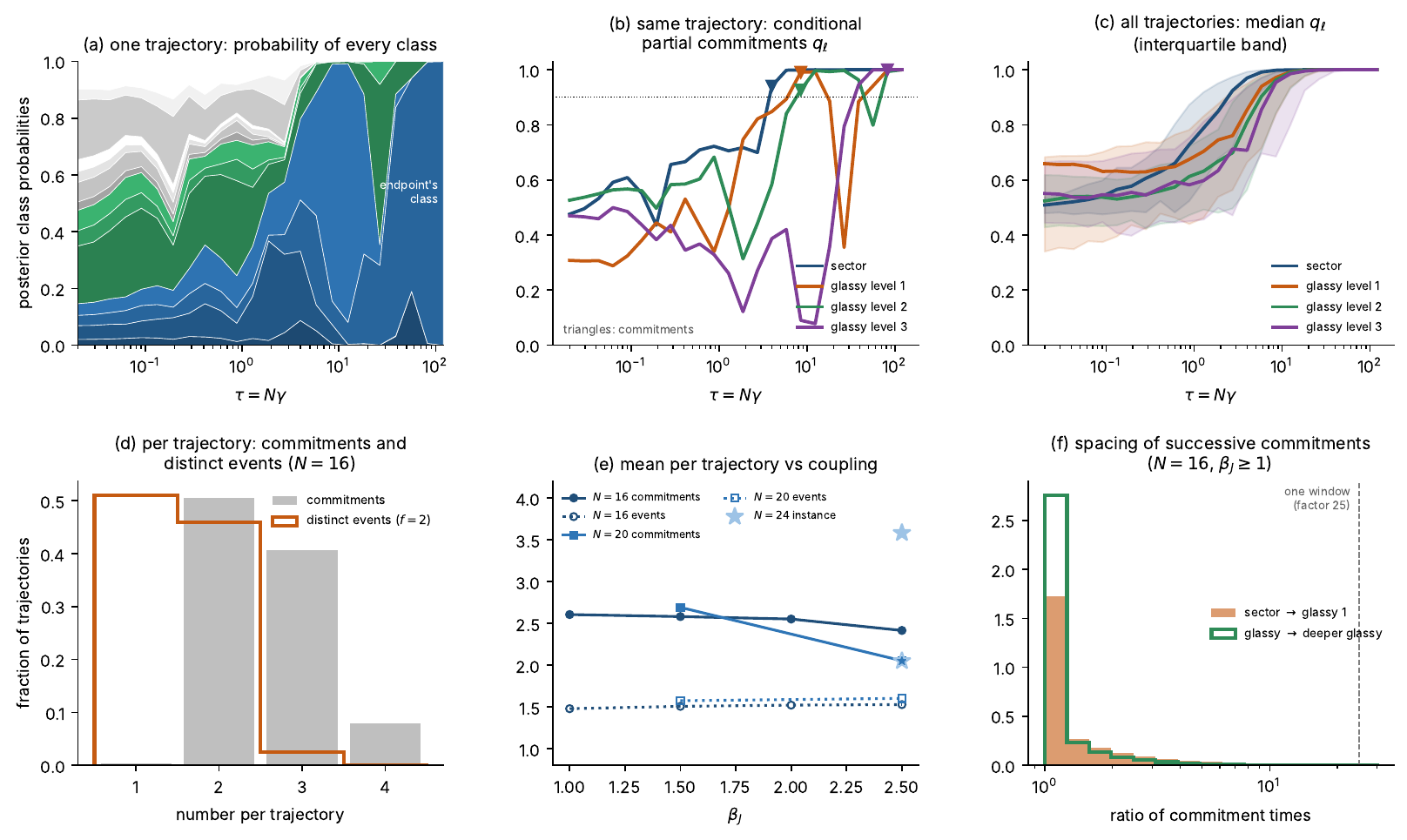}
\caption{\textbf{How a trajectory descends the tree.} (a) Partial commitments of one trajectory of the $N=24$ instance of Fig.~\ref{fig:anatomy} to
all $14$ finest classes, stacked; blue and green: the two glassy classes of the chosen sector and their subclasses; grey: the classes of the other
sector. (b) The same trajectory's conditional partial commitments $q_\ell=p_{c_\ell(Y)}/p_{c_{\ell-1}(Y)}$ along the endpoint's branch; triangles:
commitments ($p_\ell$ first reaches $0.9$); dotted: $0.9$. (c) Median and interquartile band of $q_\ell$ over the trajectories of the instance whose
branch splits at level $\ell$. (d) Distribution of the number of
commitments and of distinct events per trajectory, $N=16$ Bethe glasses with glassy levels at $\beta_J\ge1$. (e) Mean numbers of commitments and events
per trajectory against coupling, at $N=16$ and $20$, and for the $N=24$ instance (stars). (f) Distribution of the ratio of successive commitment times;
dashed: the factor $25$ spanned by one decision window.}
\label{fig:partial}
\end{figure}

\emph{Question.} The ensemble decides the levels at different times. Does a single trajectory then make one decision, one decision per level, or
something in between? \emph{Data.} For the instance of Fig.~\ref{fig:anatomy}, the full distribution over its $14$ finest classes along every
trajectory. For counting, all $N=16$ Bethe instances at $\beta_J=1$, $1.5$, $2$ and $2.5$ ($240$ instance--temperature pairs, of which $133$ have glassy
levels; $128$ trajectories each) and the $12$ $N=20$ instances at $\beta_J=1.5$ and $2.5$ ($64$ trajectories each), with posteriors on $40$ values of
$\gamma$. Fig.~\ref{fig:partial} shows one trajectory in detail (panels a and b), all trajectories of the instance (panel c), and the counts over
many instances (panels d--f).

\emph{One trajectory (panels a and b of Fig.~\ref{fig:partial}).} Panel a shows the partial commitments of one trajectory to all $14$ finest classes, stacked so that the bands
add up to one. Grey bands are the classes of the other sector; green and blue bands are the two glassy classes of the chosen sector and their
subclasses; the endpoint's finest class $c_L(Y)$ is marked. Up to $\tau\approx1$ the bands keep roughly their equilibrium widths: the
trajectory has not yet decided anything. The grey bands then drain away, and the trajectory commits to its sector at $\tau\approx4$. Next the green
bands drain: the trajectory commits to its first and second glassy classes together, at $\tau\approx8.5$, and shortly afterwards recrosses at the first
glassy level before committing again. Finally, around $\tau\approx80$, the weight inside the chosen class concentrates on the endpoint's finest class (third
glassy level). Panel b shows the same trajectory through its conditional partial commitments $q_\ell=p_\ell/p_{\ell-1}$
\eqref{eq:branchcommit}, the probability of the endpoint's class $c_\ell(Y)$ given its parent $c_{\ell-1}(Y)$; the triangles mark the commitments,
the first times at which $p_\ell$ reaches $0.9$. The descent is explicit here: the sector's $q$ rises first, the $q$'s of
the first two glassy levels rise together, the first of them dips during the recrossing, and the third glassy level is decided last.

\emph{All trajectories of the instance (panel c of Fig.~\ref{fig:partial}).} The panel shows the median of each $q_\ell$ over the trajectories whose branch splits at level $\ell$, with the interquartile
range shaded. The medians complete half of their rise at $\tau\approx1.3$ for the sector, at $\tau\approx4$ for the first two glassy levels and at
$\tau\approx6$ for the third. \emph{The highlighted trajectory is typical: a sector decision followed by a compound glassy decision.}

\emph{Counting over many instances (panels d--f of Fig.~\ref{fig:partial}).} A trajectory \emph{commits} at level $\ell$ when $p_\ell$ first exceeds $0.9$, and commitments within a
factor two of each other in $\gamma$ are counted as one \emph{distinct event} (Sec.~\ref{sec:partial}). Panel d shows the distribution of the number of
commitments (grey bars) and of distinct events (orange outline) per trajectory, for $N=16$ instances with glassy levels at $\beta_J\ge1$. A trajectory
commits $2.55$ times on average---the sector plus one to three glassy levels---but in only $1.50$ distinct events: one compound event in $51\%$ of the
trajectories, two (the sector, then the glassy levels together) in $46\%$, and three or more in $2\%$. Panel e shows the mean numbers of commitments
and events against coupling: both are nearly independent of $\beta_J$; at $N=20$ the number of events is slightly larger ($1.57$--$1.60$), and in the
$N=24$ instance (stars) a trajectory commits $3.58$ times in $2.05$ events. Panel f shows the ratio of successive commitment times along a trajectory,
separately for the step from the sector to the first glassy level (orange) and for steps between glassy levels (green); the dashed line marks the
factor $25$ spanned by one decision window. Successive glassy commitments coincide (median ratio $1.0$; $70\%$ within a factor $1.5$), while the sector
and the first glassy commitment are further apart (median ratio $1.5$; $42\%$ at least a factor two). All ratios are far smaller than the width of one
window.

\emph{Size, robustness, recrossings and the ferromagnet.} The two-step descent becomes more frequent as $N$ grows: two or more events occur in $48\%$,
$55\%$ and $80\%$ of the trajectories at $N=16$, at $N=20$ and in the $N=24$ instance, and the sector and glassy commitments are at least a factor two
apart in $44\%$, $49\%$ and $67\%$ (at $\beta_J=1.5$ and $2.5$). The counts do not hinge on the thresholds: for commitment thresholds from $0.8$ to $0.95$
and merging factors from $1.5$ to $3$ the mean number of events at $N=16$ stays between $1.3$ and $1.8$. After committing, $14\%$ of the trajectories
recross at the sector and $12\%$ at the first glassy level. In the ferromagnet on the same graphs nearly every trajectory makes a single decision, the
mirror choice.

\emph{Conclusion.} A trajectory makes neither a single decision nor one decision per level. It descends the tree in one or two steps---the mirror
choice and a compound glassy decision---because only the sector's $\lambda_P$ differs substantially from the others, while the glassy parents have
similar $\lambda_P$ and hence overlapping windows (Sec.~\ref{sec:channel}).

\subsection{How the classes change with temperature}\label{sec:temperature}
\begin{figure}[!htbp]\centering
\includegraphics[width=\linewidth]{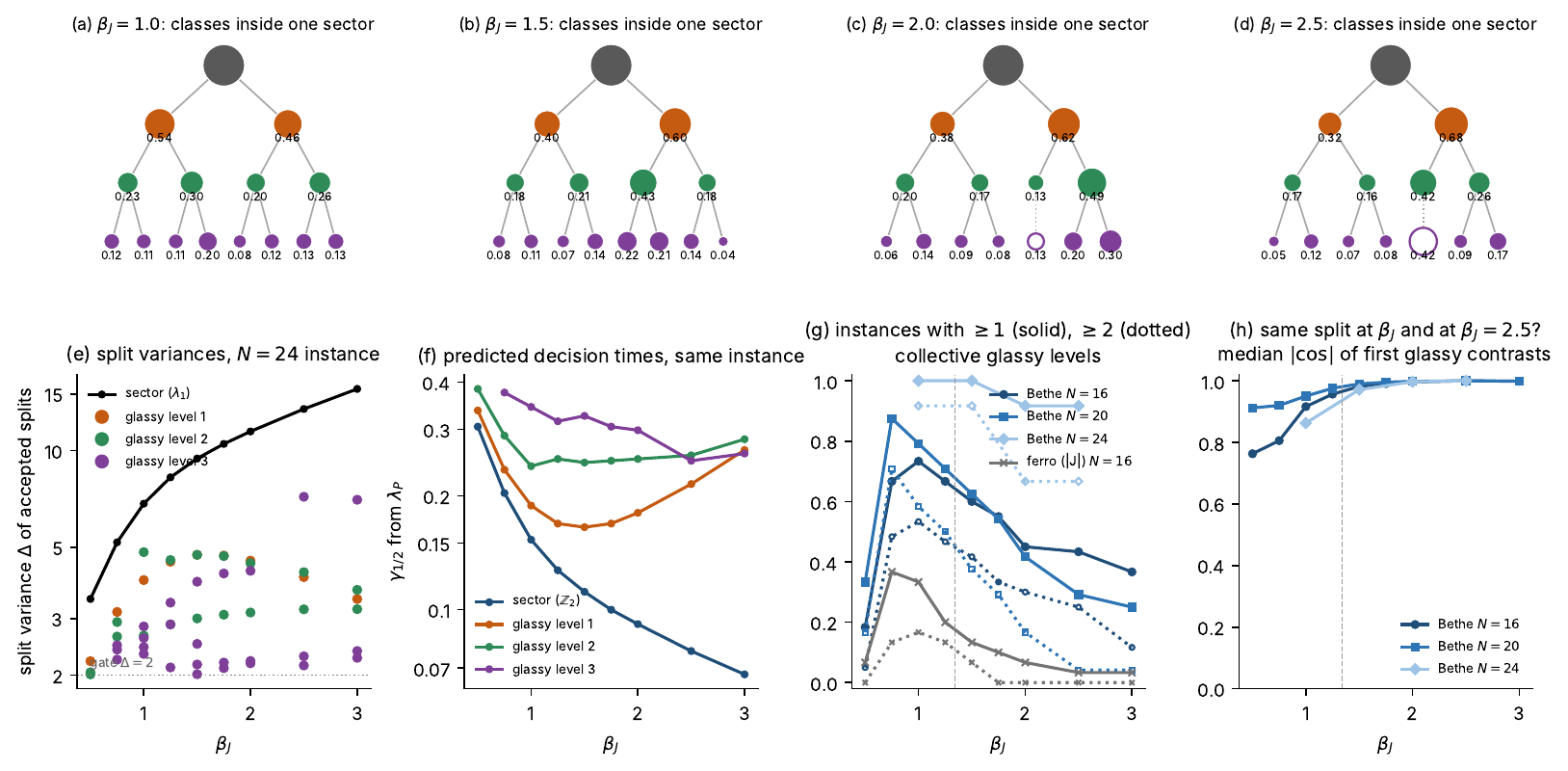}
\caption{\textbf{Classes across temperature.} (a--d) Classes inside one sector of the $N=24$ instance of Fig.~\ref{fig:anatomy} at four values of
$\beta_J$ (numbers: masses within the sector; open circles: classes not split further). (e) Split variance $\Delta$ of every accepted split of this instance, coloured by level, and the sector
eigenvalue $\lambda_1$; dotted: the collectivity gate $\Delta=2$. (f) Predicted decision times of each level of this instance. (g) Fraction of unselected
instances with at least one (solid) and at least two (dotted) collective glassy levels; grey: ferromagnet on the same graphs; dashed: Bethe transition.
(h) Median $|\cos|$ between the first glassy contrast at $\beta_J$ and at $\beta_J=2.5$, for instances that have both.}\label{fig:temperature}
\end{figure}

\emph{Question.} Are the glassy classes a property of the glass phase? And as the target is cooled, does the generator face the same classes, or new
ones? \emph{Reference temperatures.} In the thermodynamic limit the Bethe glasses studied here have their glass transition at $\beta_c\approx1.34$,
i.e.\ at temperature $T_c=1/\beta_c\approx0.75$, where $2\,\E\tanh^2(\beta_cJ^{(0)})=1$ (Sec.~\ref{sec:procedure}); the ferromagnet on the same graphs
orders at $\beta_J\approx0.83$. At $N\le24$ there is no sharp transition but a broad crossover; the dashed line in Fig.~\ref{fig:temperature}g, h
marks $\beta_c$, and the fraction of instances with glassy classes (panel g) peaks somewhat below it, at $\beta_J\approx0.75$--$1$. \emph{Data.} The classes and predicted decision times of the same disorder instances at nine values of $\beta_J$ between $0.5$ and $3$. Because the
couplings are $J=\beta_JJ^{(0)}$, the ground state and hence the alignment do not depend on $\beta_J$, and a split at one temperature can be compared
with a split at another through the $|\cos|$ between their unit contrast vectors. Fig.~\ref{fig:temperature} follows one instance across temperature
(panels a--f) and then summarizes all instances (panels g and h).

\emph{One instance (panels a--f of Fig.~\ref{fig:temperature}).} Panels a--d show the classes inside one sector of the $N=24$ instance of Fig.~\ref{fig:anatomy} at $\beta_J=1$, $1.5$,
$2$ and $2.5$ (panel d is the tree of Fig.~\ref{fig:anatomy}a). At $\beta_J=0.5$ (not shown) the procedure finds two weak glassy levels with
$\Delta\approx2$; from $\beta_J=0.75$ on it finds three. From $\beta_J=1.25$ on, the first glassy split is the same split as at $\beta_J=2.5$
($|\cos|\ge0.96$ between the contrasts; $0.83$ at $\beta_J=1$ and $0.43$ at $\beta_J=0.5$). Cooling changes the masses---the larger glassy class grows from
$0.60$ to $0.73$ of the sector between $\beta_J=1.25$ and $3$---but not the classes themselves. Panel e shows the split variances $\Delta$ of all accepted
splits, coloured by level, together with the sector's $\lambda_1$; panel f shows the predicted decision times of each level. Cooling moves the sector
decision earlier, because $\lambda_1$ grows from $3.5$ to $15.5$ and the predicted $\gamma_{1/2}$ falls from $0.31$ to $0.067$, while the glassy
decisions stay at $\gamma_{1/2}\approx0.17$--$0.28$. The sector and glassy decisions are therefore simultaneous at $\beta_J=0.75$ and a factor four apart
at $\beta_J=3$. The predicted decision time of the first glassy level (orange in panel f) is
not monotonic in $\beta_J$: it first decreases and then increases again, with a minimum near $\beta_J\approx1.5$. It is set by $\lambda_P$, the leading
eigenvalue of the correlation matrix inside a sector. On cooling, $\lambda_P$ first grows, from $3.1$ at $\beta_J=0.5$ to $6.4$ at $\beta_J=1.5$, because
the two glassy classes become more sharply distinct; it then falls, to $4.0$ at $\beta_J=3$, because the smaller class loses weight (from $0.40$ to
$0.27$ of the sector) and the fluctuations inside each class freeze, so that less variance is left within the sector for the split to explain. The
split variance of the first glassy split follows the same rise and fall ($\Delta=2.2$, $4.7$ and $3.5$ at $\beta_J=0.5$, $1.5$ and $3$; panel e). The
deeper levels have parents with smaller and less temperature-dependent $\lambda_P$, so their predicted times change little, while the sector's
$\lambda_1$ grows monotonically, so its decision moves steadily earlier.

\emph{All instances (panels g and h of Fig.~\ref{fig:temperature}).} Panel g shows the fraction of unselected instances with at least one (solid) and at least two (dotted)
collective glassy levels, at each size, together with the ferromagnet on the same graphs (grey); the dashed line marks the Bethe transition. Collective
glassy levels are rare in the paramagnet ($18\%$ of the instances at $N=16$ and $27\%$ at $N=20$ for $\beta_J=0.5$), most common near the transition
($73\%$ at $\beta_J=1$ for $N=16$; $88\%$ at $\beta_J=0.75$ for $N=20$), and less common deep in the glass phase at these sizes ($37\%$ and $25\%$ at
$\beta_J=3$). At $N=24$ they are present in $12$ of $12$ instances at $\beta_J=1$ and $1.5$, and in $11$ of $12$ at $\beta_J=2$ and $2.5$. The decrease at
low temperature reflects a shift of weight, not a disappearance of classes: the median mass of the smaller class of the best first split falls from
$0.34$ to $0.22$ ($N=16$) and from $0.36$ to $0.14$ ($N=20$) between $\beta_J=0.5$ and $3$, so that its $\Delta$ falls below the gate. Panel h shows, for
instances that have a first glassy split both at $\beta_J$ and at $\beta_J=2.5$, the median $|\cos|$ between the two contrasts. For $\beta_J\ge1$ it is
$0.92$--$1.00$, $0.95$--$1.00$ and $0.86$--$1.00$ at $N=16$, $20$ and $24$: wherever the split exists, it is the same split.

\emph{Conclusion.} At these sizes, glassy classes appear near the glass transition and keep their identity on cooling. What changes with temperature is
their weights and, through $\lambda_1$, how far the mirror choice precedes the glassy decision.

\subsection{Across instances and sizes}\label{sec:scan}
\begin{figure}[!htbp]\centering
\includegraphics[width=\linewidth]{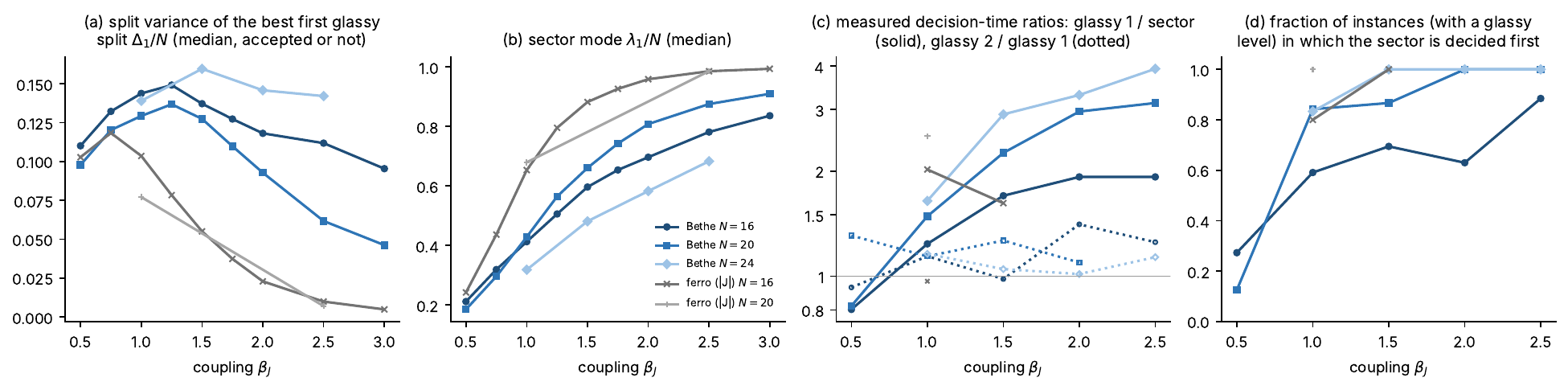}
\caption{\textbf{Across instances and sizes}, Bethe glasses and ferromagnets on the same graphs ($|J|$). (a) Median split variance per spin,
$\Delta_1/N$, of the best first glassy split, accepted or not. (b) Median sector eigenvalue per spin, $\lambda_1/N$. (c) Median measured ratios of
decision times: first glassy level over sector (solid) and second over first glassy level (dotted). (d) Fraction of instances with glassy levels in
which the sector is decided first.}\label{fig:scan}
\end{figure}

\emph{Question.} Do glassy splits involve a finite fraction of the spins, as the word ``collective'' suggests? And how does the gap between the mirror
choice and the glassy decision change with $N$? \emph{Data.} All unselected Bethe instances and the ferromagnetic controls (Fig.~\ref{fig:scan}), with
measured decision times at $N=16$ and $20$ and for six $N=24$ instances.

\emph{Collectivity (panels a and b of Fig.~\ref{fig:scan}).} Panel a shows the median split variance per spin, $\Delta_1/N$, of the best first glassy split, whether or not it
passes the gates. In the glass phase $\Delta_1/N\approx0.13$--$0.16$ at $\beta_J=1$--$1.5$ for $N=16$, $20$ and $24$, with no systematic decrease with
$N$; a split of a fixed number of spins would instead give $\Delta_1/N\propto1/N$. In the ferromagnet, $\Delta_1/N$ falls to $0.01$ on cooling. Panel b
shows the sector eigenvalue per spin, $\lambda_1/N$, which grows from about $0.2$ in the paramagnet to $0.7$--$0.9$ at $\beta_J=2.5$, and to $0.99$ in the
ferromagnet.

\emph{Separation of the decisions (panels c and d of Fig.~\ref{fig:scan}).} Panel c shows the measured ratio of the first glassy decision time to the sector decision time
(solid) and of the second glassy decision time to the first (dotted). The first ratio grows with coupling and with size: it is $1.7$, $1.9$ and $1.9$
at $N=16$, $2.3$, $3.0$ and $3.1$ at $N=20$, and $2.9$, $3.3$ and $3.9$ at $N=24$, for $\beta_J=1.5$, $2$ and $2.5$. The second ratio stays between $1.0$ and
$1.4$. Panel d shows the fraction of instances with glassy levels in which the sector is decided first: $59$--$88\%$ at $N=16$ and all instances at
$N=24$ for $\beta_J\ge1.5$.

\emph{Comparison with the prediction.} Over the same instances, the predicted ratio of first glassy to sector decision time grows from $2.3$ to $3.1$
($\beta_J=1.5$) and from $3.1$ to $4.8$ ($\beta_J=2.5$) between $N=16$ and $24$. The reason is spectral: $\lambda_1$ grows in proportion to $N$, whereas
the glassy parents' $\lambda_P$ grows only from about $3.4$ to $4.0$ ($\lambda_P/N$ falls from $0.21$ to $0.16$; whether glassy splits stay extensive as
$N\to\infty$ cannot be settled at $N\le24$). The measured ratios are $5$--$40\%$ smaller than predicted because the sector is decided somewhat later
than predicted, for a reason identified in Sec.~\ref{sec:lawexp}.

\emph{Conclusion.} At these sizes glassy splits involve a finite fraction of the spins, and the mirror choice separates from the glassy decision as
$N$ grows, as the spectra predict.

\subsection{Testing the binary-channel prediction}\label{sec:lawexp}

\begin{figure}[!htbp]\centering
\includegraphics[width=\linewidth]{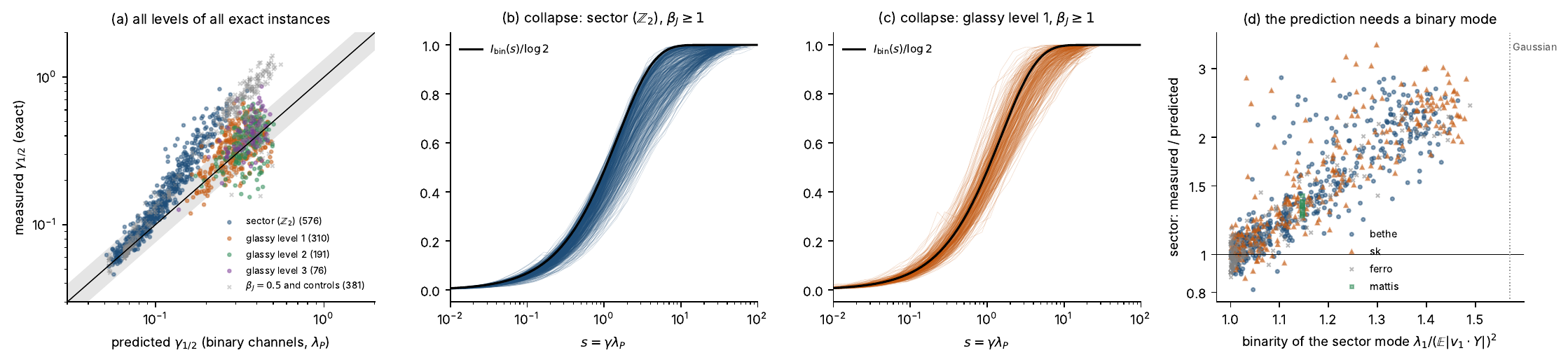}
\caption{\textbf{The binary-channel prediction.} (a) Measured against predicted decision time, for every level of every exact instance, coloured by level;
grey crosses: $\beta_J=0.5$ and the controls; band: $\pm30\%$. (b, c) Information about the sector (b) and the first glassy level (c) against the
rescaled clock $s=\gamma\lambda_P$, one curve per instance with $\beta_J\ge1$; black: $I_{\rm bin}(s)/\log2$. (d) Ratio of measured to predicted
sector decision time against the binarity index $\lambda_1/(\E|v_1\cdot Y|)^2$ of the sector mode; dotted: the Gaussian value $\pi/2$.}\label{fig:law}
\end{figure}

\emph{Question.} Can the decision times be predicted from the equilibrium measure alone, before any trajectory is generated? As explained in
Sec.~\ref{sec:channel}, the binary-channel prediction \eqref{eq:law} is an approximation built on one idealization of each split; nothing else in the
analysis depends on it. This subsection measures how accurate it is. \emph{Data.} Measured
and predicted decision times for every level of every exact instance (Fig.~\ref{fig:law}).

\emph{All levels (panel a of Fig.~\ref{fig:law}).} The panel plots the measured decision time against the predicted one, for every level of every exact instance; the grey
band marks agreement within $\pm30\%$. For the glassy levels of Bethe glasses at $\beta_J\ge1$, the ratio of measured to predicted time has median
$1.03$, $1.01$ and $1.09$ at the first, second and third glassy level (interquartile ranges within $0.87$--$1.25$; $400$ measurements), and $68$--$74\%$ of
the measurements lie within $\pm30\%$. No parameter is fitted.

\emph{Universal curves (panels b and c of Fig.~\ref{fig:law}).} The panels plot the information about the sector (b) and the first glassy level (c) of every instance with
$\beta_J\ge1$ against the rescaled clock $s=\gamma\lambda_P$, together with the binary-channel curve $I_{\rm bin}(s)/\log2$. The curves of different
instances collapse on $I_{\rm bin}$, and the collapse does not narrow between $N=16$ and $24$, as expected for a crossover of fixed width
(Sec.~\ref{sec:channel}).

\emph{Where the prediction errs (panel d of Fig.~\ref{fig:law}).} The prediction rests on two assumptions: that the split is binary, i.e.\ that the projection of the configurations on the
parent's leading eigenvector takes two sharp values, and that the split lies along that eigenvector. For the sector the first assumption can fail.
The sector is decided later than predicted (median ratio $1.24$ at $\beta_J\ge1$ and $2.3$ in the paramagnet), and panel d shows that the delay is set by
how far $v_1\cdot Y$ is from two sharp peaks: it is plotted against the binarity index $\lambda_1/(\E|v_1\cdot Y|)^2$, which equals $1$ for two sharp
peaks and $\pi/2$ for a Gaussian. The two are strongly rank-correlated (Spearman $0.90$ over $874$ instances, controls included), and the $253$ instances
with an index below $1.05$ have median ratio $1.03$. For glassy splits it is the second assumption that can fail, which matters most in SK glasses near
the transition (Sec.~\ref{sec:families}). The label-free version of the prediction, which uses the successive eigenvalues of the equilibrium correlation matrix
$C_0$ instead of the parents' $\lambda_P$, works equally well on other instances, including planar lattices at $N=32$--$64$
(Fig.~\ref{fig:legacystatic}), and on the controls.

Equilibrium spectra provide a useful leading decision-time prediction. Its limitations include sector binarity, alignment of glassy splits with their parents' leading modes, and the imbalance of the conditional child masses; the latter is tested independently in Appendix~\ref{app:additional_validation}.

\subsection{Sparse and dense glasses}\label{sec:families}
\begin{figure}[!htbp]\centering
\includegraphics[width=\linewidth]{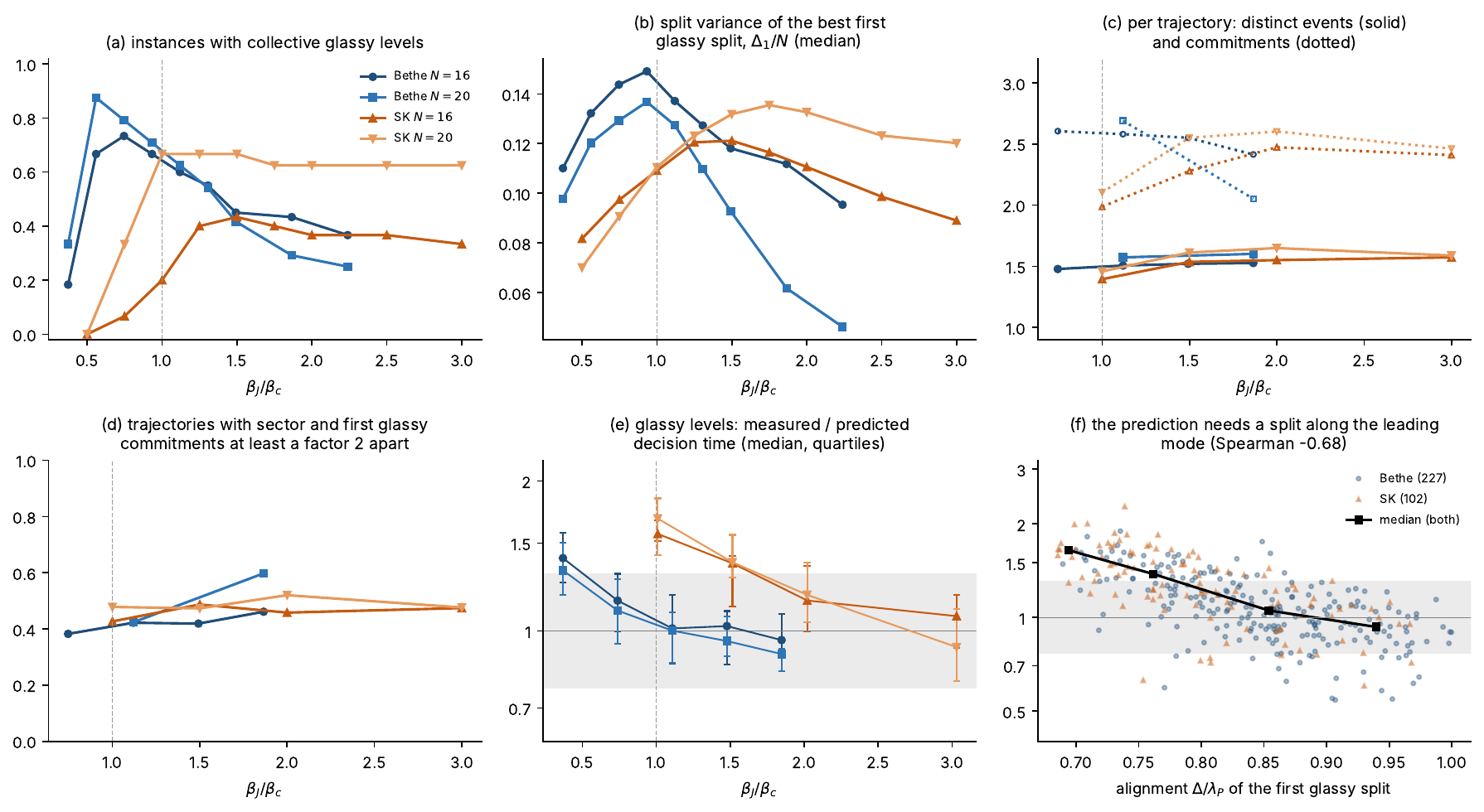}
\caption{\textbf{Sparse (Bethe) and dense (SK) glasses}, compared at equal distance from the transition, $\beta_J/\beta_c$ (dashed: $\beta_c$).
(a) Fraction of unselected instances with collective glassy levels. (b) Median split variance per spin, $\Delta_1/N$, of the best first glassy split,
accepted or not. (c) Mean numbers of distinct events (solid) and commitments (dotted) per trajectory, for instances with glassy levels. (d) Fraction of
trajectories whose sector and first glassy commitments are at least a factor two apart. (e) Ratio of measured to predicted decision time of the glassy
levels (median and quartiles); band: $\pm30\%$. (f) The same ratio for the first glassy level of every instance, against the alignment
$\Delta/\lambda_P$ of the split with the parent's leading mode; squares: binned medians.}\label{fig:families}
\end{figure}

\emph{Question.} Bethe glasses are sparse and locally tree-like; SK glasses are dense. Does this change the way trajectories commit, or only the classes
they commit to? \emph{Data.} SK glasses at $N=16$ ($30$ instances) and $N=20$ ($24$ instances), analysed exactly as the Bethe glasses: classes on the
nine-point temperature grid, and measured decision times and per-trajectory commitments at four temperatures. Fig.~\ref{fig:families} compares the two
families at equal distance from their transitions, $\beta_J/\beta_c$.

\emph{How trajectories commit: the same (panels c and d).} Panel c shows the mean numbers of distinct events (solid) and commitments (dotted) per
trajectory, for instances with glassy levels. An SK trajectory commits $2.0$--$2.6$ times in $1.4$--$1.65$ distinct events, against $2.1$--$2.7$ times in
$1.5$--$1.6$ events for Bethe glasses. Panel d shows the fraction of trajectories whose sector and first glassy commitments are at least a factor two
apart: $43$--$52\%$ for SK and $38$--$60\%$ for Bethe glasses. In both families successive glassy commitments coincide (median ratio $1.0$), and
recrossings are equally frequent ($12$--$15\%$ at the sector, $12\%$ at the first glassy level, averaged over temperature).

\emph{What they commit to: different (panels a and b of Fig.~\ref{fig:families}).} Panel a shows the fraction of instances with collective glassy levels. At $N=16$, SK glasses
have them in at most $43\%$ of the instances (at $1.5\beta_c$), against $60$--$79\%$ for Bethe glasses between $0.75\beta_c$ and $1.1\beta_c$. At $N=20$ the
SK fraction rises to $67\%$ from $\beta_c$ on and stays at $62\%$ down to $3\beta_c$, whereas in Bethe glasses it falls to $25\%$ at $2.2\beta_c$. The SK
deficit at $N=16$ is therefore largely a finite-size effect of a dense model with only $16$ spins. Panel b shows the median split variance per spin of
the best first glassy split, with the same contrast: in Bethe glasses $\Delta_1/N$ peaks near the transition and falls on cooling ($0.05$ at $2.2\beta_c$
for $N=20$), as the minority class loses weight (Sec.~\ref{sec:temperature}); in SK glasses it peaks deeper in the glass phase ($0.14$ at $1.75\beta_c$)
and stays at $0.12$ down to $3\beta_c$.

\emph{How well the prediction describes them: different near the transition (panels e and f).} Panel e shows the ratio of measured to predicted decision time
for the glassy levels. SK glassy levels are decided later than predicted near the transition (median ratio $1.6$--$1.7$ at $\beta_c$ and $1.4$ at
$1.5\beta_c$) and approach the prediction only deep in the glass phase ($0.93$--$1.07$ at $3\beta_c$); in Bethe glasses the ratio stays between $0.90$ and
$1.02$ from $\beta_c$ on. Panel f explains the difference. It plots the same ratio for the first glassy level of every instance against the alignment
$\Delta/\lambda_P\le1$ of the split with the parent's leading mode, which the prediction assumes to be one. The alignment is $0.73$--$0.79$ for SK glasses at
$\beta_c$ and $1.5\beta_c$, $0.87$--$0.89$ at $3\beta_c$, and $0.86$--$0.96$ for Bethe glasses. Across all $329$ first glassy splits of both families, the ratio
falls as the alignment improves (Spearman $-0.68$): its median is $1.38$ for alignments between $0.7$ and $0.8$, $1.05$ between $0.8$ and $0.9$, and
$0.93$ above $0.9$.

\emph{Conclusion.} Sparse and dense glasses differ in how often and how strongly glassy classes exist, and in how closely their splits follow the
leading mode of their parents. However, they do not differ in how trajectories commit. This is what Sec.~\ref{sec:decisions} implies: a trajectory sees the
target only through its classes and their spectra.

\subsection{Does the protocol matter?}\label{sec:protocol}
\emph{Question.} For the ideal, continuous-time generator the protocol $\nu(t)$ does not change what is decided, only when (Sec.~\ref{sec:pid}). A
generator that takes finitely many steps, however, places its steps on the $\gamma$ axis according to the protocol. Does it then matter which protocol
is used, and in which part of the path? \emph{Data.} Nine $N=16$ Bethe instances with glassy levels (three at $\beta_J=1.5$, six at $\beta_J=2.5$). The
generator \eqref{eq:sde} is integrated with $K=2$ to $64$ Euler steps, uniformly spaced in $t$, using the \emph{exact} drift \eqref{eq:drift}; the
endpoint is $y=\operatorname{sign}(X_1)$; and $4000$ trajectories are generated per protocol and $K$ ($2000$ at $\beta_J=1.5$). We compare three harmonic
protocols, which differ in how they distribute physical time over the clock $\gamma$: $\nu=0$, which spreads the decisions over roughly the first $40\%$
of $t$ and leaves half of $t$ to the \emph{local stage} $\gamma\ge1$, in which individual spins are fixed; a constant $\nu$ tuned by the prediction so that the
predicted decisions fall at $t=1/2$ (median $\nu=13$), which leaves a quarter of $t$ to the local stage; and a stiff $\nu=144$, which leaves an eighth.

\begin{figure}[!htbp]\centering
\includegraphics[width=\linewidth]{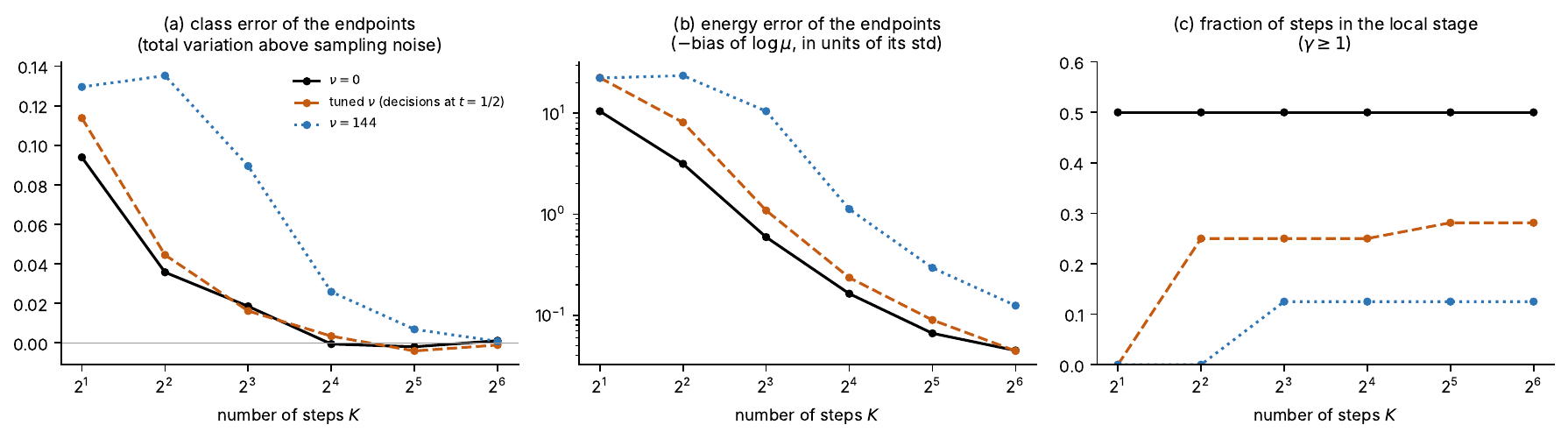}
\caption{\textbf{Protocol and discretization}, nine $N=16$ Bethe glasses, exact drift, three harmonic protocols, against the number $K$ of Euler steps
(medians over instances). (a) Class error of the endpoints: total-variation distance of the finest-class frequencies from the exact masses, minus the
sampling-noise value. (b) Energy error: shortfall of the mean $\log\mu$ of the endpoints, in units of its equilibrium standard deviation. (c) Fraction
of the $K$ steps in the local stage $\gamma\ge1$.}\label{fig:protocol}
\end{figure}

\emph{What is measured (Fig.~\ref{fig:protocol}).} Panel a shows the class error of the endpoints: the total-variation distance between the frequencies
of the finest classes among the generated endpoints and their exact masses, minus the distance expected from sampling noise alone (computed from the
same number of exact samples). Panel b shows the energy error: the amount by which the mean of $\log\mu(y)$ over the endpoints falls short of its
equilibrium value, in units of its equilibrium standard deviation; a positive value means that the endpoints have too high an energy. Panel c shows the
fraction of the $K$ steps that fall in the local stage $\gamma\ge1$ under each protocol.

\emph{Findings.} Both errors fall as $K$ grows and vanish within sampling noise for $K\ge32$. At small $K$, $\nu=0$ is the best protocol and $\nu=144$
the worst, at both couplings. At $K=4$ the class error above sampling noise is $0.036$ for $\nu=0$, $0.044$ for the tuned protocol and $0.135$ for $\nu=144$;
at $K=8$ it is $0.018$, $0.016$ and $0.089$, and the endpoints' $\log\mu$ falls short by $0.6$, $1.1$ and $10.5$ standard deviations. The ranking follows the
share of steps in the local stage (panel c), not the coverage of the decision windows: the tuned protocol puts more steps into the sector window than
$\nu=0$ ($40\%$ against $25\%$) and still does worse. Even the class errors arise mainly in the local stage, because the endpoint is read off the final
spins. With three protocols the two kinds of coverage cannot be fully separated, but none of the data favour moving steps from the local stage to the
decision windows.

\emph{Conclusion.} For a discretized generator with an exact drift, the macroscopic decisions are robust to coarse steps---we attribute this to the drift
being the exact posterior mean---and steps are needed where individual spins are fixed. A learned, approximate drift may shift this balance.

\emph{A practical recommendation.} For a generator with an exact (or very accurate) drift and $K$ steps uniformly spaced in $t$, our data suggest
choosing the protocol so that about half of the steps fall in the local stage, i.e.\ so that $\gamma(t)$ reaches one near $t=1/2$. Among harmonic
protocols this is $\nu=0$, for which $\gamma(1/2)=1$ exactly; with it, $K=16$ steps already bring both errors to the level of sampling noise (class
error within sampling noise, energy shortfall $0.16$ standard deviations), whereas the tuned protocol needs about $32$ steps and $\nu=144$ about $64$.
Equivalently, if a stiffer protocol is preferred for other reasons, the steps should not be uniform in $t$ but placed so that about half of them fall
where $\gamma\ge1$. The decision windows themselves need only a few steps: at $K=4$ the class error of $\nu=0$ is already below $0.04$. This
recommendation is a tentative extrapolation from three protocols, nine $N=16$ instances and an exact drift. We have not tested protocols with an even
larger share of steps in the local stage (for example $\nu<0$), and with a learned drift, whose errors are largest where the posterior is most
uncertain, steps in the decision windows may become more valuable.

\subsection{Negative controls}\label{sec:controls}
\begin{figure}[!htbp]\centering
\includegraphics[width=\linewidth]{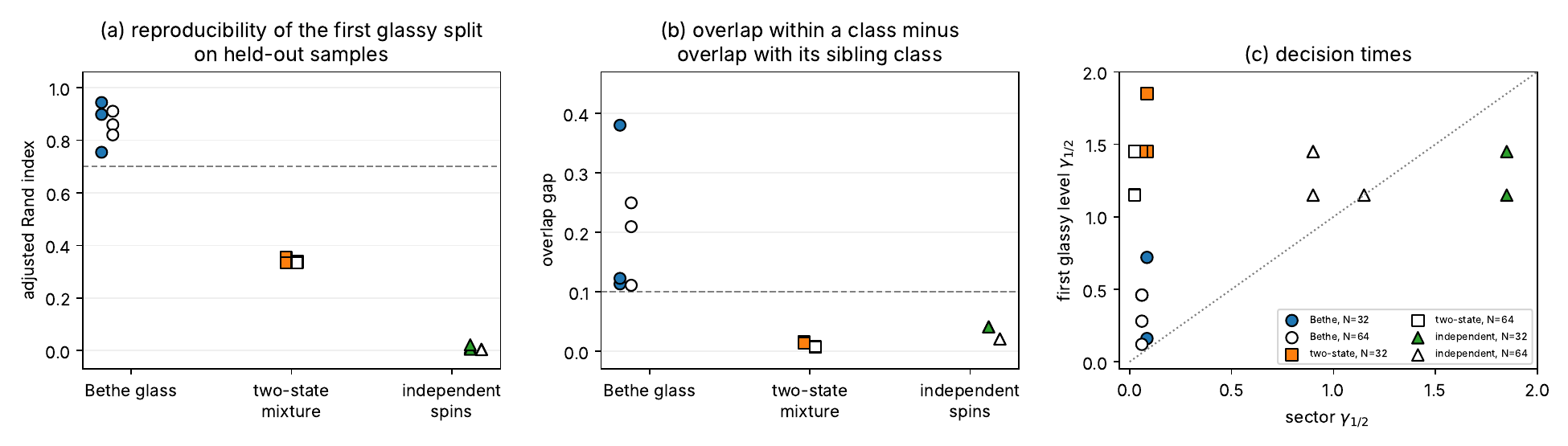}
\caption{\textbf{Structural tests at $N=32$ and $64$}, Bethe glasses (three instances per size) and two controls (filled: $N=32$; open: $N=64$).
(a) Reproducibility of the first split below the sector: adjusted Rand index between partitions found on independent sample sets. (b) Mean overlap
within a class minus mean overlap with its sibling class. Dashed: acceptance thresholds. (c) Decision time of the first split below the sector against
that of the sector; dotted: equal times.}\label{fig:controls}
\end{figure}

\emph{Question.} Could the procedure produce the same picture for a target without glassy structure? \emph{Data.} The ferromagnet on the same graphs and
the Mattis model at $N=16$ (exact), and independent spins and a two-state mixture at $N=32$ and $64$ (sampled, Fig.~\ref{fig:controls}); the controls are
described in Sec.~\ref{sec:procedure}.

\emph{Exact controls at $N=16$.} The ferromagnet on the same graphs has the same sector as the glass, with an even larger $\lambda_1$, but its splits
inside a sector disappear on cooling. Its $\Delta_1/N$ falls from $0.10$ at $\beta_J\le1$ to $0.01$ at $\beta_J=2.5$ (Fig.~\ref{fig:scan}a), and the
fraction of instances with a glassy level falls from $10/30$ to $1/30$ at $N=16$, and from $3/12$ to $0/12$ at $N=20$ (Fig.~\ref{fig:temperature}g); its
trajectories make one decision, the mirror choice. The Mattis model never passes the collectivity gate: its best split inside a sector has
$\Delta_1=0.81$ in all ten instances.

\emph{Sampled controls at $N=32$ and $64$ (Fig.~\ref{fig:controls}).} Here the classes are identified from equilibrium samples, and two structural tests
decide whether a split below the sector is genuine. Panel a shows its reproducibility, measured by the adjusted Rand index between the partitions
found on two independent sets of samples ($1$: identical; $0$: unrelated); panel b shows the mean overlap of configurations within a class minus their
mean overlap with the sibling class. The dashed lines are the acceptance thresholds ($0.7$ and $0.1$). The Bethe glasses pass both tests; independent
spins and the two-state mixture fail both. Panel c shows why timing alone is not enough: the two-state mixture also has a fast sector decision followed
by a much later ``first glassy'' one, but the latter is a split of thermal noise, which panels a and b reject. An unfrustrated mixture of four product
states with a built-in two-level tree, by contrast, passes the tests in four of five replicas (not shown).

\emph{Conclusion.} The procedure does not invent glassy classes: frustration is what keeps collective splits inside a sector in the ordered phase. Panel c
illustrates consequence (4) of Sec.~\ref{sec:channel}: separated decision windows alone do not demonstrate a hierarchy.

\subsection{Larger systems and learned generators}\label{sec:larger}\label{sec:learned}
\emph{Question.} Does the sequence---mirror choice first, then the glassy decision---persist beyond exact enumeration? And does a trained denoiser
represent the same decisions?

\begin{figure}[!htbp]\centering
\includegraphics[width=\linewidth]{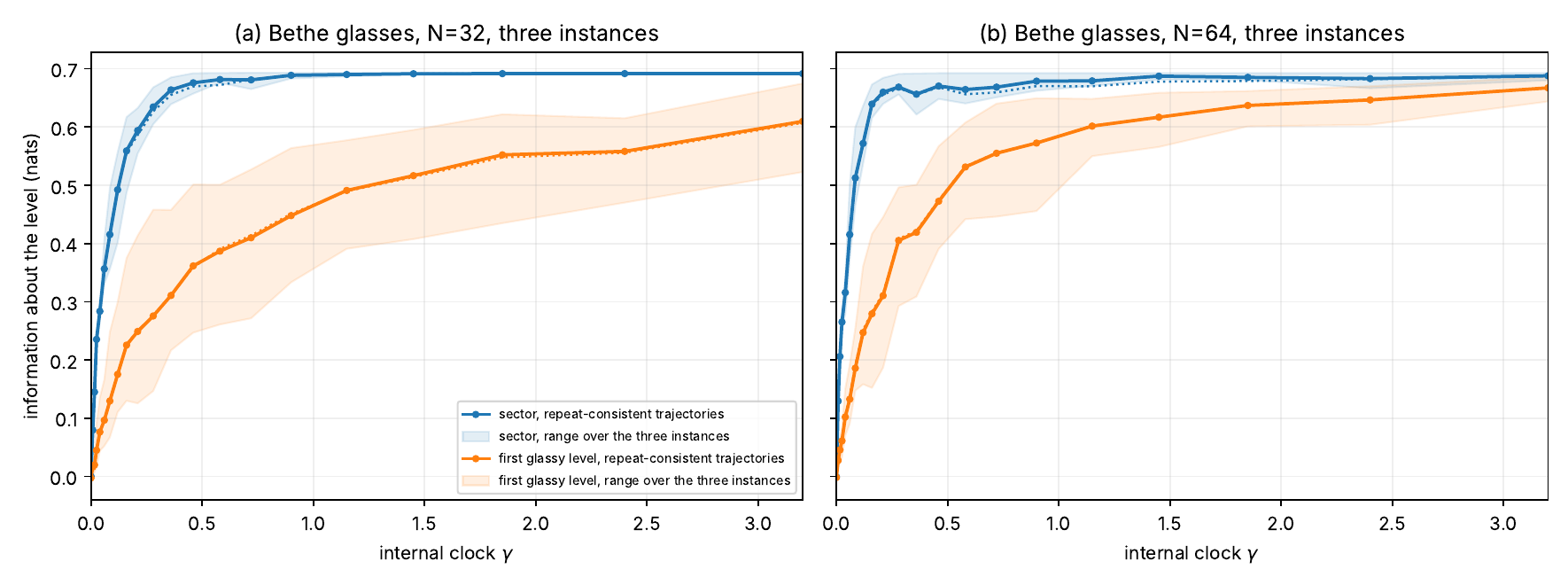}
\caption{\textbf{Larger systems}, three Bethe instances each at $N=32$ (a) and $N=64$ (b), sampled posteriors. Information about the sector (blue) and
the first glassy level (orange) against the internal clock $\gamma$; solid: average over repeat-consistent trajectories; dotted: all trajectories;
shading: range over the three instances.}\label{fig:larger}
\end{figure}

\emph{Larger systems (Fig.~\ref{fig:larger}).} For three Bethe instances at each of $N=32$ and $64$, the posteriors along $24$ trajectories per
instance are estimated by sequential Monte Carlo on $21$ values of $\gamma$ up to $3.2$ (Sec.~\ref{sec:procedure} and Appendix~\ref{app:methods}).
A trajectory is called \emph{repeat-consistent} when two independent Monte Carlo runs give the same class posteriors along it (total-variation
difference at most $0.1$ at every time); $71$ of $72$ trajectories per size are. The figure shows the information about the sector and about the first
glassy level against $\gamma$, averaged over the repeat-consistent trajectories (solid) and over all trajectories (dotted), with the range over the
three instances shaded. The sector is decided first, already at the first grid points ($\gamma_{1/2}=0.060$--$0.085$, an upper bound; the predicted value is
$0.02$--$0.06$), and the first glassy level later, at $\gamma_{1/2}=0.12$--$0.72$. All $71$ repeat-consistent trajectories per size commit to a sector, and
$67$ ($N=32$) and $70$ ($N=64$) commit to a glassy class; $8$ and $6$ recross at the sector, and $7$ and $9$ at the glassy level. The sequence thus persists
at these sizes; because the instances were selected, they say nothing about how common glassy classes are.

\begin{figure}[!htbp]\centering
\includegraphics[width=\linewidth]{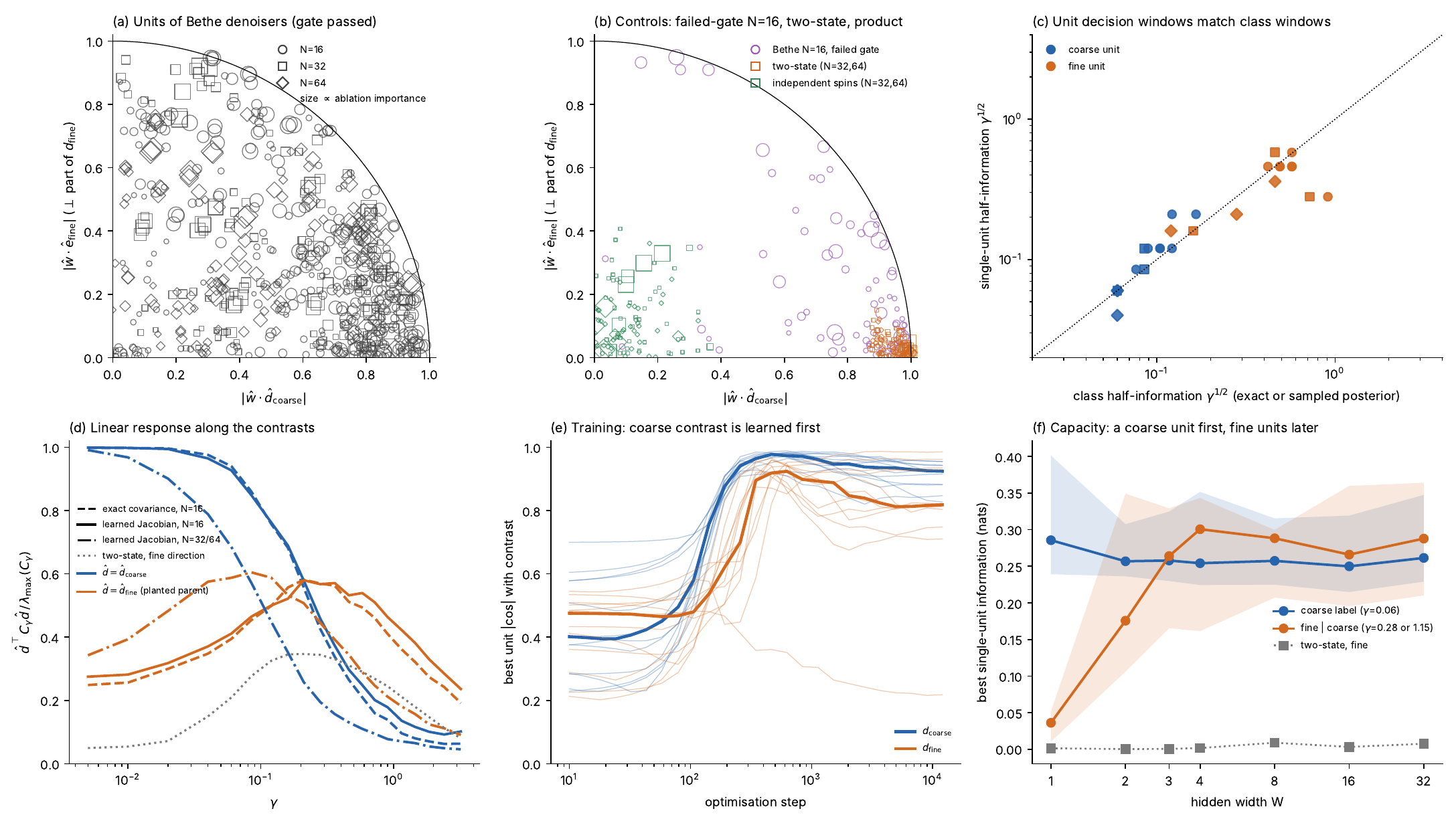}
\caption{\textbf{Hidden units of learned denoisers} (``coarse'': sector contrast; ``fine'': first glassy contrast). (a) Alignment of each hidden
unit's input weights with the sector contrast (horizontal) and with the orthogonal part of the glassy contrast (vertical), Bethe instances whose splits
pass the gates; symbol size: importance. (b) The same for controls. (c) Decision time of the most informative unit against the exact decision time of
the class. (d) Fraction of the leading posterior fluctuation along each contrast, $\hat d^\top C_\gamma\hat d/\lambda_{\max}(C_\gamma)$, for the exact
covariance and the learned Jacobian. (e) Best unit--contrast alignment during training. (f) Information of the best single unit against the width of
the hidden layer.}\label{fig:learned}
\end{figure}

\emph{Learned generators (Fig.~\ref{fig:learned}).} A denoiser trained only on pairs $(Y,H_\gamma)$ learns an approximation of the posterior mean $\yh$,
and its Jacobian with respect to the field approximates the posterior covariance $C_\gamma$ (Eq.~\eqref{eq:known}). We trained small one-hidden-layer
denoisers for twelve $N=16$ and six $N=32,64$ instances (Appendix~\ref{app:methods}) and asked whether their hidden units encode the class contrasts.
Panel a places each hidden unit of the Bethe denoisers according to the alignment of its input weights $\hat w$ with the sector contrast (horizontal)
and with the part of the first glassy contrast orthogonal to it (vertical); units on the arc lie in the plane of the two contrasts, and symbol size is
proportional to the unit's importance (the increase of the test error when the unit is replaced by its mean). Many important units lie near the arc:
$58\%$ of the importance-weighted weight is in the plane of the two contrasts, against at most $12.5\%$ for random directions. Panel b shows the same
for controls: units of the two-state denoisers align with the sector contrast only, units of the independent-spin denoisers with neither. Panel c
compares the decision time of the most informative unit for each level---the clock value at which the unit carries half of its final information
about the level---with the exact decision time of the level: they agree within factors $1.15$ (sector) and $1.24$ (glassy level). Panel d follows,
along $\gamma$, how much of the leading fluctuation of the exact covariance (dashed) and of the learned Jacobian (solid and dash-dotted) lies along
each contrast: in both, the sector contrast dominates at small $\gamma$ and the glassy contrast takes over around $\gamma\approx0.2$, the hand-over of
Fig.~\ref{fig:anatomy}d (median overlap of the leading eigenvectors $0.98$). Panel e shows that during training the sector contrast is learned first and
the glassy contrast later (final $|\cos|=0.92$ and $0.82$). Panel f shows the information carried by the best single unit against the width $W$ of the
hidden layer: one unit learns the sector only; from two units on the glassy contrast is learned as well; no width yields a glassy unit for the
two-state control. Memorization of the training set affects only large $\gamma$: denoisers trained on $64$ or $1024$ configurations reproduce the
leading spectrum and the decision windows, and their Jacobian susceptibility is inflated $15$- to $297$-fold only at $\gamma=10$
(Fig.~\ref{fig:legacystatic}e, f).

\emph{Conclusion.} The sequence persists beyond exact enumeration, and trained denoisers learn the same contrasts, in the same order, as the exact
posterior. These small networks on small targets are a laboratory in which the right features are known exactly, not a model of large generators.

\FloatBarrier
\section{Discussion}\label{sec:discussion}
\paragraph{What the paper shows.} On the theory side (Secs.~\ref{sec:pid} and \ref{sec:decisions}), PID from a point source is stochastic
localization run on a protocol-dependent clock: for an ideal generator the protocol decides when its decisions happen and what they cost, but not
what is decided. The state of the generator is described by its partial commitments to the classes of the target, their average gives the
information per level, and the decision time of each split is predicted, approximately, by the leading eigenvalue of its parent. On the
experimental side (Sec.~\ref{sec:experiments}), finite Ising glasses are descended in two stages: the mirror choice, which every ordered Ising model
makes, followed by a compound decision among glassy classes, which exist near and below the glass transition and not in the ferromagnet on the same
graphs (Secs.~\ref{sec:anatomy}--\ref{sec:temperature} and \ref{sec:controls}). The two stages separate as coupling and size grow
(Sec.~\ref{sec:scan}), their timing follows the parents' spectra (Sec.~\ref{sec:lawexp}), the per-trajectory picture is the same in sparse and dense
glasses (Sec.~\ref{sec:families}), and a generator with finitely many steps is limited mainly by the local stage (Sec.~\ref{sec:protocol}).

\paragraph{Other sampling applications.} Nothing in Secs.~\ref{sec:pid}--\ref{sec:levels} is specific to spins except the exact enumeration. The
same questions can therefore be asked of diffusion-type generators in the other areas where they are now used for sampling
\citep{yang2023diffusion,carleo2019ml}: equilibrium ensembles of molecules \citep{noe2019boltzmann,noe2020mlmolsim}, molecular design
\citep{alakhdar2024diffusion}, crystalline materials \citep{debreuck2025crystal,zeni2025mattergen}, lattice field theory \citep{cranmer2023lattice},
and turbulent flows and other physical fields \citep{li2024lagrangian}. There the posteriors must be estimated by Monte Carlo
\citep{delmoral2006smc} or read off a trained denoiser (Sec.~\ref{sec:learned}). Six lessons of this study carry over, each established above for
spin glasses.
\begin{enumerate}[label=(\roman*),leftmargin=*]
\item \emph{Remove symmetries first.} A symmetry of the target, like spin inversion, forces a decision that every trajectory must make and that
says nothing about the structure of interest. For molecules, global rotations, permutations of identical atoms and mirror images play this role, and
configurations should be aligned before classes are defined, as in Sec.~\ref{sec:defs}.

\item \emph{What is decided belongs to the target.} The protocol sets only when the decisions happen (Sec.~\ref{sec:clock}) and, for a discretized
generator, by how accurately the ideal one is approximated. We oserve that with an accurate drift the step budget is best spent in the local stage, by the rule
given in Sec.~\ref{sec:protocol}. Whether this holds for learned drifts is directly testable in these applications.

\item \emph{Spectra and conditional masses, not tree depth alone, order commitments.} In an approximately balanced binary regime, covariance eigenvalues predict the timing of collective choices. For strongly unequal basins, however, conditional masses and centroid separation must be considered together with spectra. In molecular applications the relevant collective coordinates could include dihedral angles or metastable basin magnetizations. Classes with weak distinguishability can be decided late; however, child mass, centroid separation and within-class fluctuations all affect decision times, so mass alone does not order commitments or predict the loss of rare modes under score error.

\item \emph{Only collective rearrangements are decided early} (consequence (1)). Splits whose variance is comparable with that of a single
coordinate are decided in the local stage, together with bond lengths and other local detail.
\item \emph{Timing alone does not reveal structure} (consequence (4); Sec.~\ref{sec:controls}). Decision windows where the prediction puts them, or
well separated from each other, show only that the target has collective modes with those eigenvalues. Nested classes are shown only by the classes
themselves and the partial commitments to them.

\item \emph{The Jacobian is a label-free diagnostic.} The leading eigenvector of a trained denoiser's Jacobian along the path shows which contrast
the generator is currently deciding (Sec.~\ref{sec:learned}), and the calibration of its partial commitments, property (a) of
Sec.~\ref{sec:partial}, tests the denoiser.
\end{enumerate}

\paragraph{Limits, and why the exact study stops at $N=24$.} All statements are at $N\le24$ (exact) or $N\le64$ (selected instances, sampled
posteriors); the gates are conventional, the classes operational, and nothing is claimed about replica-symmetry breaking, ultrametricity or
$N\to\infty$. Beyond $N\approx24$ neither classes nor posteriors can be computed exactly, and a few larger instances---which we have, and which show
the same sequence---cannot decide what larger sizes would address: whether $\lambda_P/N$ of glassy splits stays finite, which would make glassy
decisions macroscopic and separated on $\tau=N\gamma$. That needs a systematic size series with approximate posteriors.

\paragraph{Outlook.} \emph{Hierarchical and compositional targets.} Images and molecules are probably less nested and more compositional than
spin glasses: largely independent local parts, each decided on its own. In a finite-dimensional glass such as the two-dimensional
Edwards--Anderson model \citep{edwards1975theory}, or in a lattice field, classes become spatial domains that may be decided in parallel rather than
in sequence; which of the lessons above survive in that case is the natural next question, and answering it requires lattices far beyond exact
enumeration. \emph{Design.} Proposition~\ref{prop:sl}(iii) turns hierarchy-aware schedules into anisotropic confinements that raise the
signal-to-noise ratio along chosen contrasts first. \emph{Larger systems.} Posteriors beyond enumeration---Monte Carlo, belief propagation and loop
corrections \citep{mezard2009information,chertkov2006loopPRE,chertkov2026book}, or trained denoisers checked by the calibration test of
Appendix~\ref{app:tools}---and a test of replica ultrametricity on posterior replicas would connect the finite-size picture to the asymptotic
theory \citep{rammal1986ultrametricity,panchenko2013ultrametricity}. 
\emph{Score geometry and information flow.}
The foam picture of Garnier-Brun \emph{et al.}\
\citep{GarnierBrun2026Foam} offers a complementary
geometric description of generative diffusion.
Their effective potential is the negative logarithm
of the marginal density of the Gaussian observation
$Z_\gamma=H_\gamma/\gamma$, whose score and Hessian
are determined by the posterior mean and covariance.
While the foam geometry reflects the full posterior
covariance, our level-resolved information production
depends on its between-class component.
An interesting question is whether the interfaces
between competing posterior states coincide with
regions of enhanced information production along
localization trajectories, and how this correspondence
depends on unresolved fluctuations within the classes.

\section*{Code and data availability}
All code, archived results, figure scripts and notebooks are available in the public repository
\url{https://github.com/mchertkov/SpinGlassHierarchy}. Every figure of the paper is regenerated from the archived results in a few minutes;
the full recomputation (exact enumeration, posterior sampling and training, several hours on two cores) is documented there in execution order
(Appendix~\ref{app:repro}).

\section*{Acknowledgments}

The author thanks Marc M\'ezard for drawing attention to Ref.~\citep{GarnierBrun2026Foam}, sharing his note on conditional overlap concentration and suggesting a direct overlap-moment diagnostic for finite-size cells.

\paragraph{Disclaimer on the use of language models.} A large language model was used to proof-read the manuscript and to help design, implement
and check the numerical experiments. All results were reviewed by the author, who is solely responsible for the content, the claims and any errors.

\appendix
\section{Proofs}\label{app:proofs}
\paragraph{Bridge coefficients.} (See \cite{behjoo2025hpid,chertkov2026adapid} for details.) For the reference process (Brownian motion with killing rate $\tfrac12\nu(t)|x|^2$) the backward kernel to
$y$ solves $\partial_t\phi+\tfrac12\nabla^2\phi-\tfrac12\nu|x|^2\phi=0$ with $\phi\to\delta(x-y)$ as $t\to1$. The ansatz
$\log\phi=-\tfrac12c|x|^2+\kappa\,x\cdot y+f(t,y)$ gives, from the $|x|^2$ and $x\cdot y$ terms, $c'=c^2-\nu$ and $\kappa'=c\kappa$; the
terminal delta requires $c(1-t)\to1$ and $\kappa(1-t)\to1$. The bridge drift is $\nabla\log\phi=\kappa y-cx$.

The conditional mean $a(t)y$ and covariance
$\sigma^2(t)I$ obey
\(
a'=\kappa-ca,\qquad
(\sigma^2)'=1-2c\sigma^2
\).
Since $\kappa'=c\kappa$, it follows that
\(
(\kappa a)'=\kappa^2,\qquad
(\kappa^2\sigma^2)'=\kappa^2
\).
Both quantities vanish initially, hence
\(
\kappa a=\kappa^2\sigma^2
=\int_0^t\kappa^2(s)\,ds=\gamma(t)
\).
This establishes the bridge coefficients
in~\eqref{eq:bridge}.

\paragraph{Constant confinement.}
For $\nu(t)\equiv S^2\geq0$, the Riccati equations with the bridge
terminal condition yield
\begin{equation}\label{eq:hpid}
\begin{gathered}
c(t)=S\coth[S(1-t)],\qquad
\kappa(t)=\frac{S}{\sinh[S(1-t)]},\\
a(t)=\frac{\sinh(St)}{\sinh S},\qquad
\sigma^2(t)=\frac{\sinh(St)\sinh[S(1-t)]}{S\sinh S},\\
\gamma(t)=\frac{S\sinh(St)}{\sinh S\sinh[S(1-t)]}.
\end{gathered}
\end{equation}
Their $S\to0$ limits are $a=t$, $\sigma^2=t(1-t)$,
$\kappa=c=(1-t)^{-1}$ and $\gamma=t/(1-t)$.
For large $S$ with $St\gg1$ and $S(1-t)\gg1$,
$\gamma(t)\simeq2S e^{-2S(1-t)}$; consequently a macroscopic
decision at $N\gamma=\tau_*$ occurs approximately at the time $t_0$
satisfying $2S(1-t_0)\simeq\log N+\log(2S/\tau_*)$.

As a process-level check, the conditioned field $h_t=\kappa(t)X_t$ satisfies
$dh_t=\kappa^2(t)y\,dt+\kappa(t)\,dW_t$,
and hence $h_t=\gamma(t)y+B_{\gamma(t)}$.

\paragraph{Proposition~\ref{prop:sl}.} (i) follows from the last display; under the mixture law the drift of $h$ is $\kappa^2\yh$, and the
innovation $\bar B$ is Brownian in the observation filtration \citep{liptser2001statistics}. (ii) With $\kappa=\sqrt{\gamma'}$ and
$X=H_{\gamma(t)}/\kappa$, It\^o's formula gives $dX=(\gamma'/\kappa)\yh\,dt-(\kappa'/\kappa)X\,dt+(\sqrt{\gamma'}/\kappa)\,dW=(\kappa\yh-cX)\,dt+dW$.
Defining $\nu=c^2-c'$, the pair $(c,\kappa)$ solves the Riccati system above, so $\kappa y-cx$ is the bridge drift of the reference process with
confinement $\tfrac12\nu|x|^2$, and $\kappa\yh-cx$ its posterior average, i.e.\ the optimal PID drift. The condition $(1-t)^2\gamma'\to1$
gives $\kappa(1-t)\to1$, $\gamma(1-t)\to1$, hence $a=\gamma/\kappa\to1$ and $\sigma^2=\gamma/\kappa^2\to0$: $X_1=Y$. (iii) For a quadratic
confinement the reference kernel is Gaussian, so the bridge is $\mathcal N(A_ty,\Sigma_t)$ and Bayes' rule gives the stated tilt; the diagonal of
$\Gamma_t$ is constant on $\{\pm1\}^N$ and drops out.

\paragraph{Proposition~\ref{prop:energy}.} The bridge drift is $b_Y=\kappa Y-cX$ and $u=\E[b_Y\mid X]$, so
$\E\|b_Y\|^2-\E\|u\|^2=\kappa^2\E\|Y-\yh\|^2=2\,d\I(Y;X_s)/ds$ by I--MMSE and $\gamma'=\kappa^2$. With $X\mid Y\sim\mathcal N(aY,\sigma^2I)$ and
$\kappa-ca=a'$, $\E\|b_Y\|^2=N(a'^2+c^2\sigma^2)$. Integrate from $0$ with $\I(Y;X_0)=0$.

\paragraph{Eq.~\eqref{eq:known}.} Summing the filtering equation $dP(y)=P(y)(y-\yh)\cdot d\bar B$ against $y$ gives $d\yh=C\,d\bar B$;
It\^o's formula gives the rate of $\E\|\yh\|^2$; I--MMSE and $\E\|Y-\yh\|^2=N-\E\|\yh\|^2$ give the third relation.

\paragraph{Proposition~\ref{prop:levelrate}.} Since $C\to Y\to H$, $\I(C;H)=\I(Y;H)-\I(Y;H\mid C)$. Conditioned on $C$ the channel is still
Gaussian with independent noise, so both terms obey I--MMSE and
$\tfrac{d}{d\gamma}\I(C;H)=\tfrac12\E\|\E[Y|H,C]-\E[Y|H]\|^2$ by orthogonality of conditional expectations. Subtracting the identities for
$C_\ell$ and $C_{\ell-1}$ and using $\sigma(C_{\ell-1})\subset\sigma(C_\ell)$ gives \eqref{eq:levelrate}. It\^o's formula for
$-\sum_cp_c\log p_c$ with \eqref{eq:filter} shows that its drift is minus the contrast term, which is the compensator statement. The budget
is the law of total covariance applied to $\mathrm{Cov}(Y\mid H)$ along the nested labels, combined with \eqref{eq:known}.

\paragraph{Equation~\eqref{eq:condprior}.} $\I(C_\ell;H\mid C_{\ell-1})=\sum_P\omega_P\,\I(C_\ell;H\mid C_{\ell-1}=P)$ and, given $C_{\ell-1}=P$,
$Y$ is distributed as $\mu(\cdot\mid P)$ while the channel is unchanged.

\FloatBarrier
\section{Positive control with a known tree}\label{app:E0}
\begin{figure}[!htbp]\centering\includegraphics[width=\linewidth]{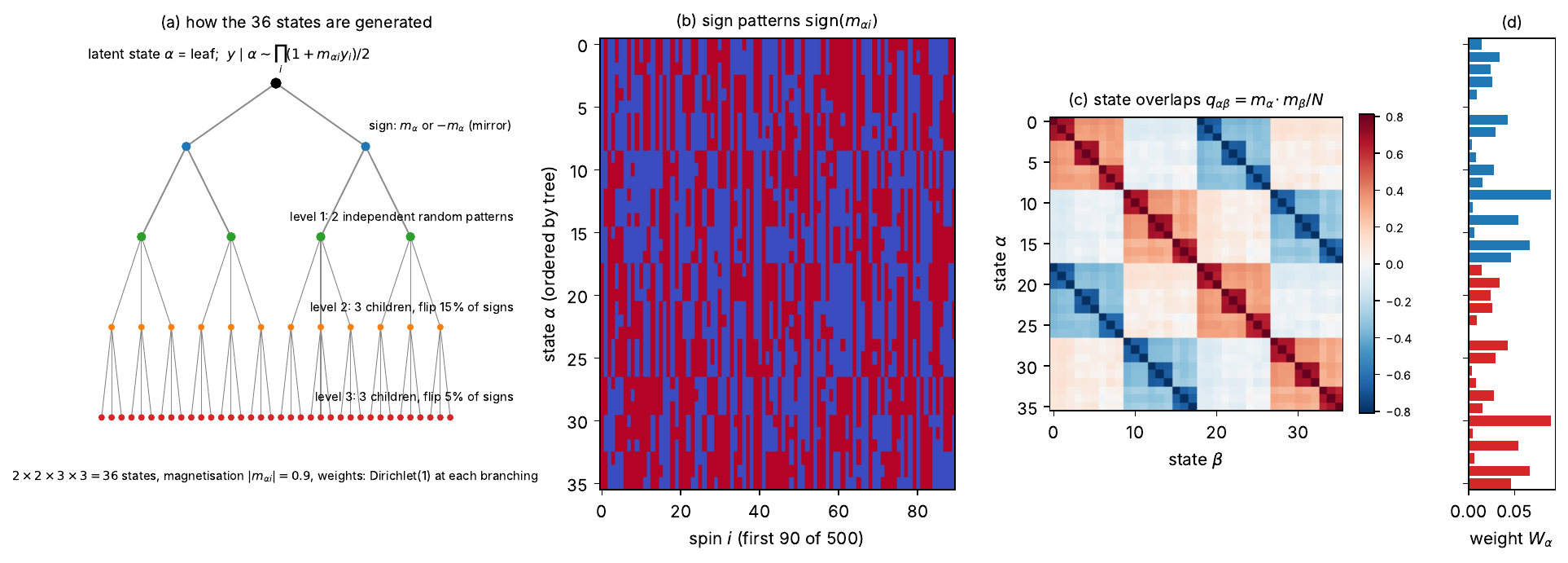}
\caption{\textbf{Positive control E0.} (a) Construction of the $36$ states on a prescribed tree, with mirror states. (b) Sign patterns on the
first $90$ spins. (c) Overlaps between states. (d) State weights.}\label{fig:E0model}
\end{figure}

\begin{figure}[!htbp]\centering\includegraphics[width=\linewidth]{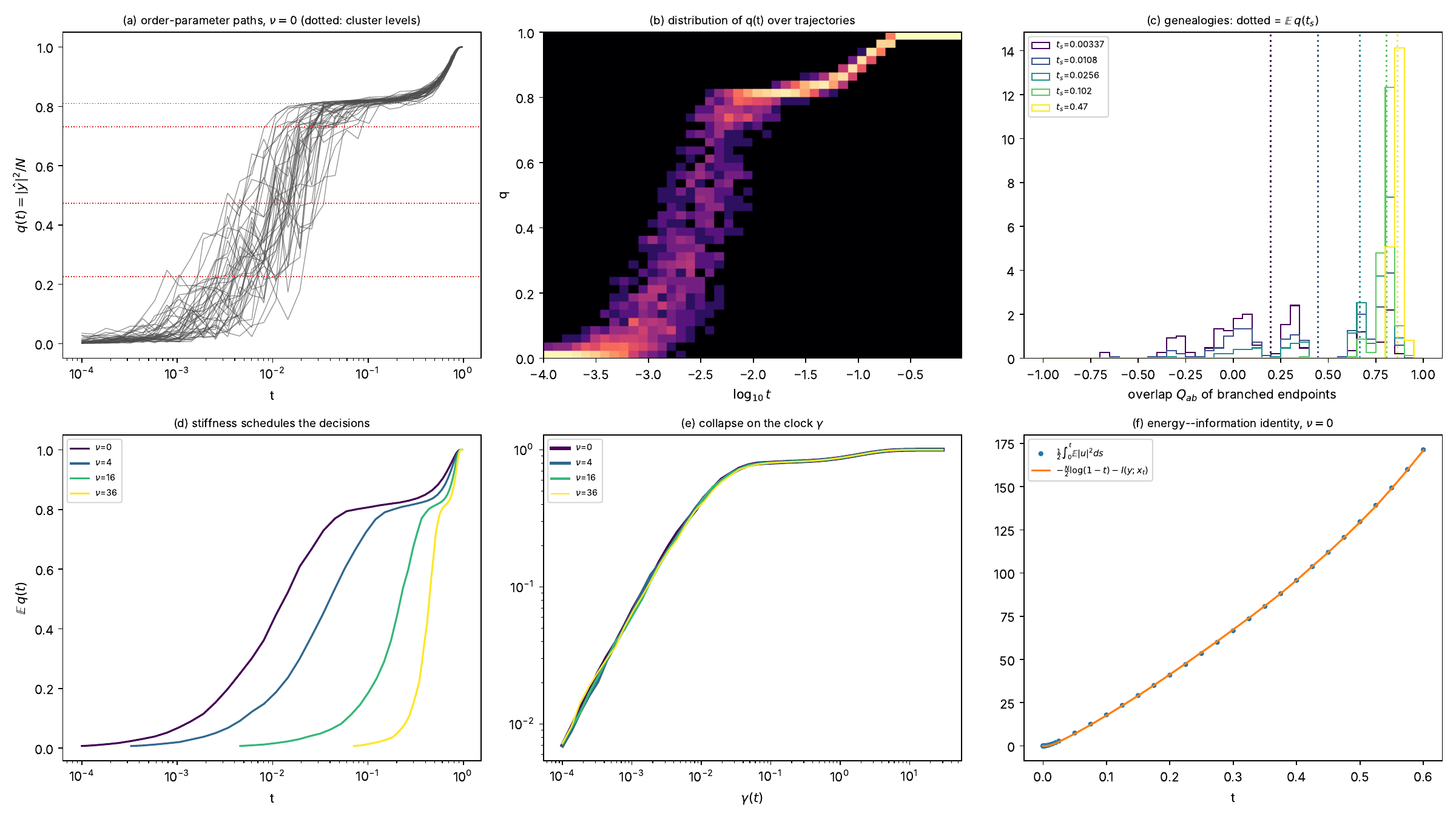}
\caption{\textbf{Identities on E0} ($N=500$). (a) Self-overlaps $q_t$ at $\nu=0$; (b) their distribution over paths; (c) overlaps of endpoints
forked at $t_s$ against $\bar q(t_s)$; (d, e) mean self-overlap for four protocols in physical time and against $\gamma$; (f) control energy
against \eqref{eq:energy}.}\label{fig:E0}
\end{figure}

Before looking for decisions in a glass the pipeline was checked on a target whose hierarchy is known by construction: a mixture of $K=36$
product measures on $N=500$ spins, $\mu(y)=\sum_\alpha W_\alpha\prod_i(1+m_{\alpha i}y_i)/2$ (Fig.~\ref{fig:E0model}). The patterns are built on a
tree: two independent level-one sign patterns, children that copy their parent and flip $15\%$ (level two) and $5\%$ (level three) of the
signs, branchings $2,3,3$, every pattern also present with reversed sign, $|m_{\alpha i}|=0.9$, Dirichlet weights. Overlaps between states are
about $0$, $0.32$, $0.66$ and $0.81$ for different level-one branches, level-two siblings, level-three siblings and within a state. The
posterior is exact (a softmax over states times product measures). With $300$ paths and $\nu=0,4,16,36$ (Fig.~\ref{fig:E0}):
the mean self-overlap collapses exactly on $\gamma$ for all protocols (Proposition~\ref{prop:sl}); endpoints of trajectories forked at
$t_s$ have mean overlaps $0.218,0.457,0.667,0.807,0.864$ against $\bar q(t_s)=0.197,0.446,0.667,0.807,0.864$; the control energy agrees with
\eqref{eq:energy} ($171.3$ against $171.0$ at $t=0.6$); median commitment times for the mirror family and the three levels are
$t=0.011,0.014,0.019,0.045$ at $\nu=0$ and $0.42,0.44,0.49,0.54$ at $\nu=36$, i.e.\ the same values of $\gamma$. The deepest level is decided
about four times later than the first, in line with the overlap gaps of the tree. E0 has no interactions; it verifies identities and pipeline,
not the emergence of a hierarchy.


\FloatBarrier
\section{Contrast identities as numerical tools}\label{app:tools}
\begin{figure}[!htbp]\centering\includegraphics[width=\linewidth]{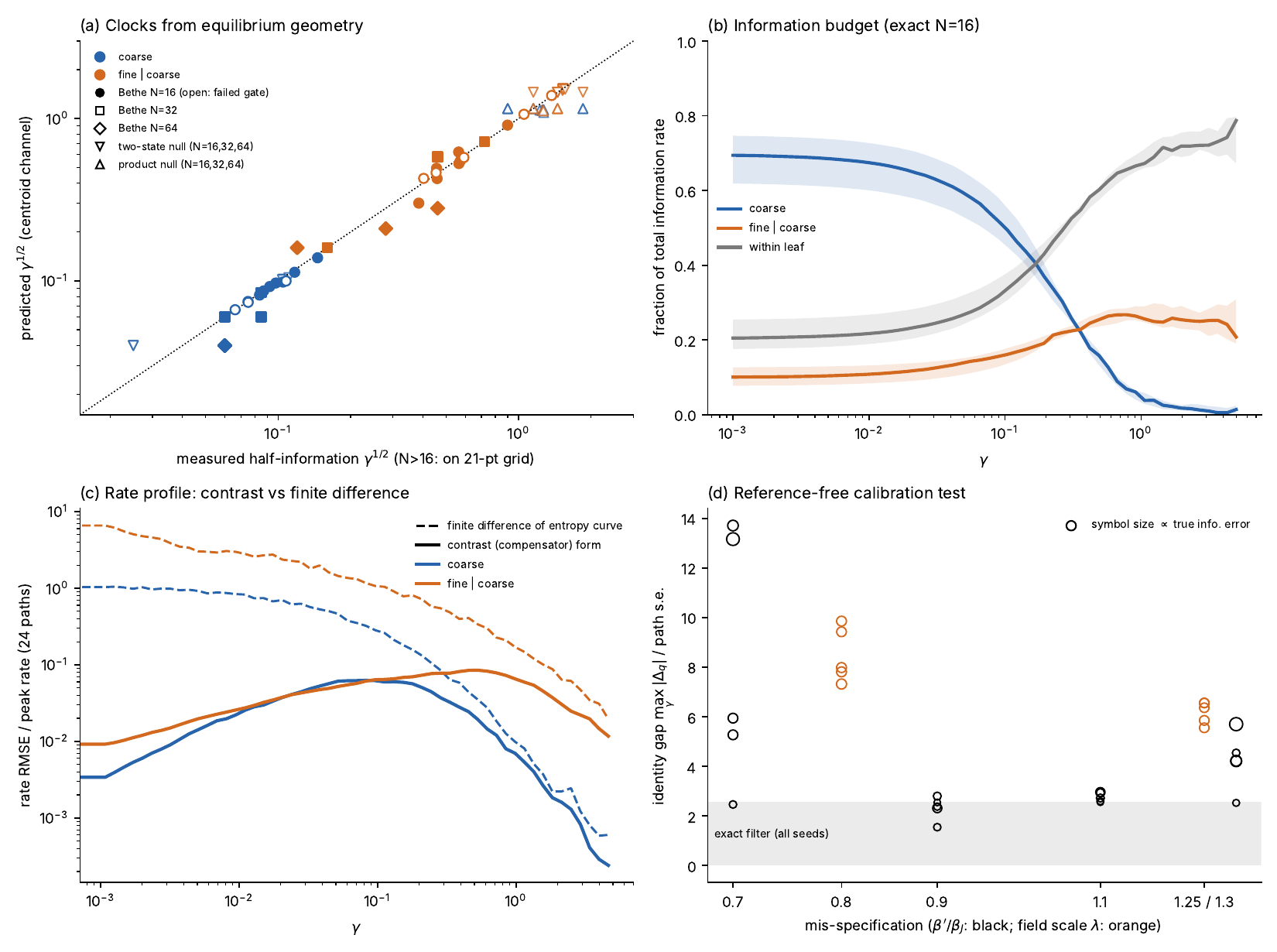}
\caption{\textbf{Contrast identities as tools} ($N=16$ instances with exact posteriors and $N=32,64$ instances with sampled posteriors). (a) Centroid-channel prediction of
half-information times; ``coarse'' is the sector, ``fine'' the first glassy split. (b) Information budget by level. (c) Rate-profile error
from $24$ paths, finite differences (dashed) versus compensator (solid). (d) Calibration gap of mis-specified posteriors versus their true
error.}\label{fig:tools}
\end{figure}

These results were obtained on twelve Bethe instances at $N=16$, $\beta_J=2.5$, with the sector and first glassy split identified from
equilibrium samples, and on ten instances of the control models, with $1536$ paths on $57$ logarithmic grid points (Fig.~\ref{fig:tools}).
\emph{Identity check.} Entropy-drop curves and integrated contrast rates agree for all $44$ level curves (largest discrepancy $0.038$\,nats;
median per-curve maximum $1.9$ path standard errors). \emph{Rate profiles.} With $24$ paths, finite differences of the entropy curve have a root-mean-square
error larger than the compensator (contrast) estimator by a median factor $6.2$ (sector) and $21.5$ (first glassy split). \emph{Window
prediction.} Replacing each leaf by its centroid gives a finite softmax channel with the same initial rates; it predicts the exact
half-information times to a median $1.7\%$ (sector) and $5\%$ (glassy split) at $N=16$, and within one grid step in $11$ of $12$ cases at
$N=32,64$---for glassy and non-glassy targets alike. \emph{Budget.} The budget of Proposition~\ref{prop:levelrate} holds to $5\times10^{-15}$; the two levels carry a
median $80\%$ of the information rate as $\gamma\to0$ and $4\%$ (sector) at $\gamma\simeq1$, where $67\%$ goes into deciding configurations
within a leaf. \emph{Reference-free calibration.} For a candidate posterior $q$, the gap between its entropy drop and its integrated contrast
rate vanishes for the Bayes filter; with $|z|>4$ as threshold it flags $18$ of $30$ deliberately mis-specified posteriors (all field-scale
errors and most $\pm30\%$ temperature errors); those that escape have class-information errors below $0.047$\,nats.

\begin{figure}[!htbp]\centering\includegraphics[width=\linewidth]{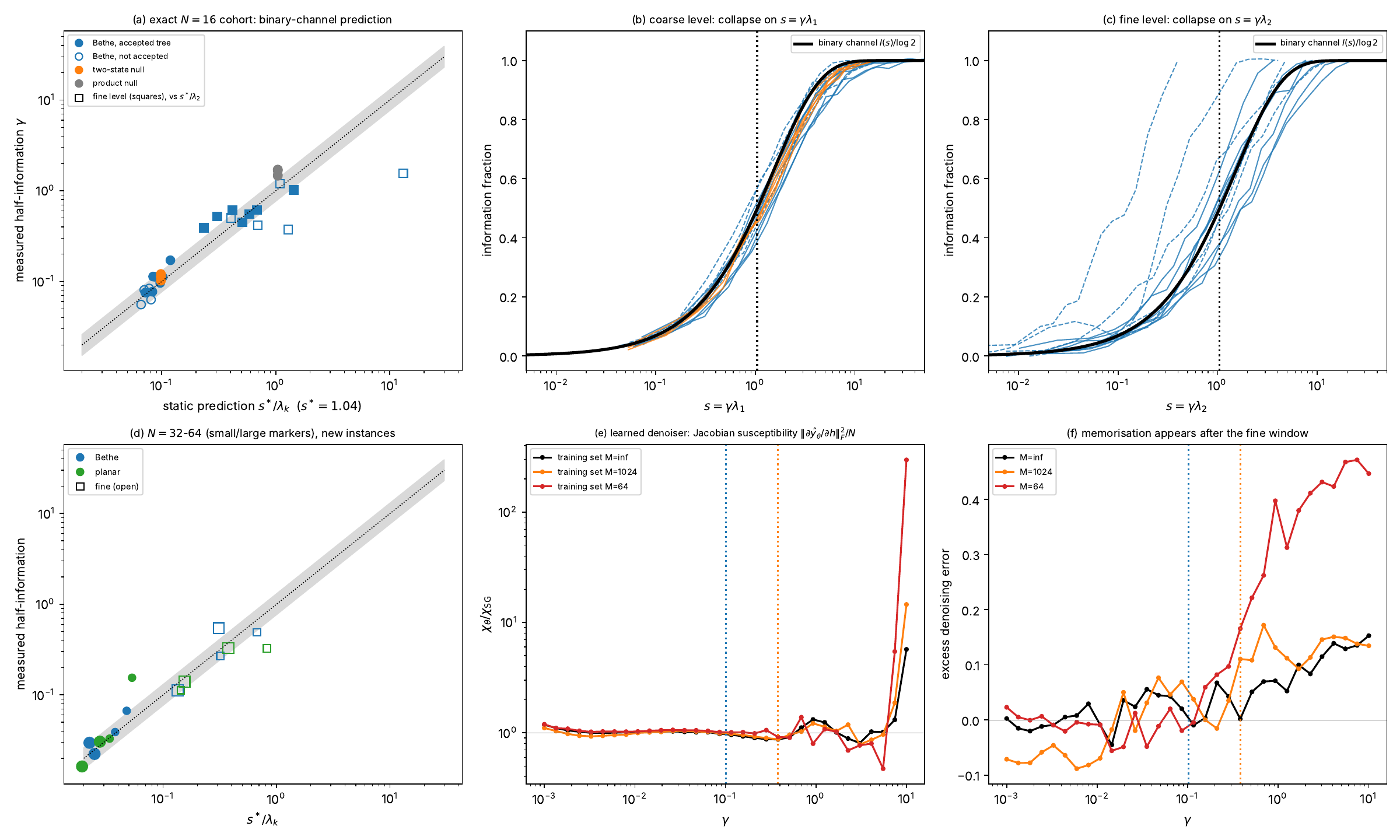}
\caption{\textbf{Label-free spectral prediction} with the successive eigenvalues $\lambda_1,\lambda_2$ of $C_0$ (``coarse'' = sector, ``fine'' =
first glassy split identified from equilibrium samples). (a) Measured half-information times of twelve $N=16$ Bethe instances against $s^*/\lambda_k$ ($k=1,2$), with
two-state and independent-spin nulls; band $\pm30\%$. (b, c) Collapse of sector and glassy information on $s=\gamma\lambda_k$. (d) Eight further
instances at $N=32$--$64$ (random regular and planar). (e) Jacobian susceptibility of learned denoisers over the exact value for three training-set
sizes. (f) Excess held-out denoising error of the same models.}\label{fig:legacystatic}
\end{figure}

\section{Additional finite-cell and mass-aware diagnostics}\label{app:additional_validation}

\begin{figure}[!htbp]\centering
\includegraphics[width=\linewidth]{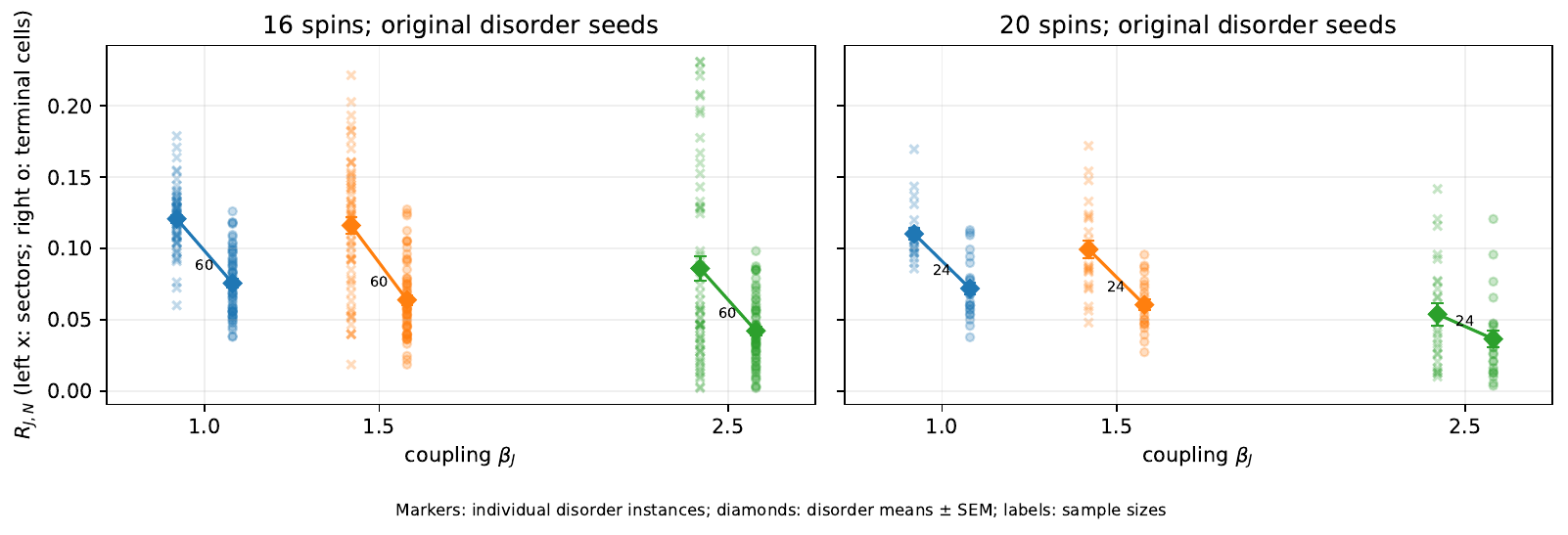}
\caption{\textbf{Conditional overlap-variance diagnostic.} Exact residual $R_{J,N}$ from the original Bethe disorder seeds, comparing the sectors-only partition (crosses, slightly left of each coupling value) with the original terminal class partition (circles, right). Diamonds and error bars show mean $\pm$ standard error over disorder; numbers give instances per coupling and size. No new class acceptance rule is used.}\label{fig:overlap_residual}
\end{figure}

\begin{figure}[!htbp]\centering
\includegraphics[width=.72\linewidth]{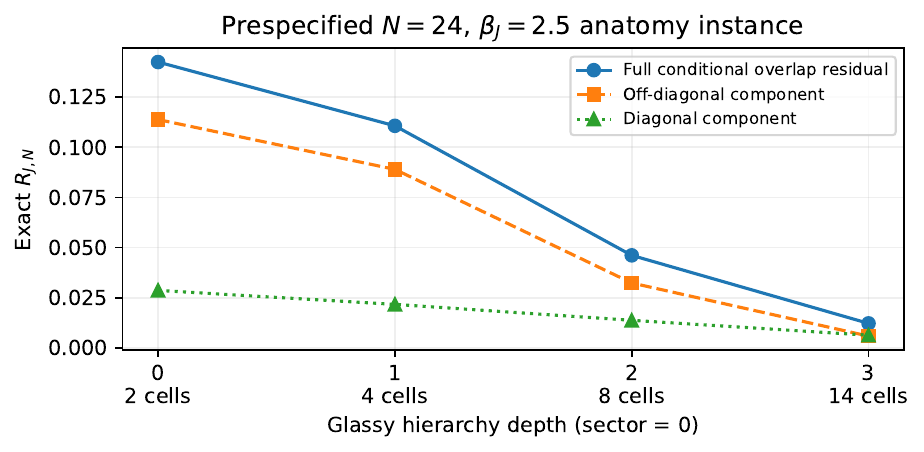}
\caption{\textbf{Overlap variance across the original $N=24$ anatomy hierarchy.} The full $R_{J,N}$ and its diagonal and off-diagonal contributions at depths $0$--$3$, with $2$, $4$, $8$, and $14$ cells. All values are exact Gibbs averages for the single prespecified disorder instance of Fig.~\ref{fig:anatomy}; the plot does not depict a disorder average.}\label{fig:anatomy_overlap}
\end{figure}

\begin{figure}[!htbp]\centering
\includegraphics[width=.87\linewidth]{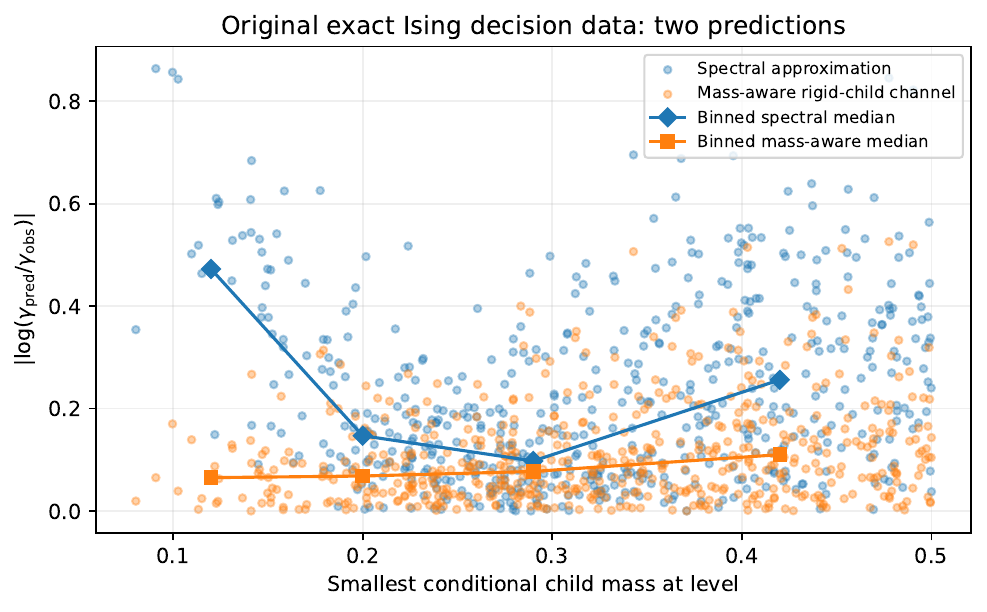}
\caption{\textbf{Unequal-mass binary-channel diagnostic.} Absolute log-ratio of predicted and observed half-information times for accepted glassy levels of the archived exact Bethe/SK instances, compared to the original parent-eigenvalue estimate. Small markers indicate individual level measurements; larger connected symbols give medians in bins of the smallest child mass, for bins with at least five observations. These are two imperfect predictions, not independent simulation replications.}\label{fig:mass_compare}
\end{figure}

Here we extend exploration of the class construction on the Bethe-glass disorder instances discussed above and analyze overlap function and related diagnostics. The terminal classes and the initial spin-inversion sectors are converted into full partitions, and Eq.~\eqref{eq:overlap_matrix} is evaluated by exact Gibbs enumeration, without posterior Monte Carlo. The results are reported in Figs.~\ref{fig:overlap_residual},\ref{fig:anatomy_overlap},\ref{fig:mass_compare}.

Fig.~\ref{fig:overlap_residual} contrasts the overlap residual for the sector-only and terminal partitions. A refinement can reduce $R_{J,N}$ while leaving nontrivial residual variance, particularly when the three-level depth cap has been reached. The data are a test of the finite-$N$ partition, not a thermodynamic extrapolation.

For every glassy level with measured half-information time archived from the exact trajectory campaign, we also compared Eq.~\eqref{eq:law} with an \emph{unequal-mass, rigid-centroid} channel. For each accepted split $P$, its child masses $w_P,1-w_P$ and observed centroid split variance $\Delta_P$ determine $d_P^2=\Delta_P/[w_P(1-w_P)]$. The predicted normalized information curve for a given level is the sum of $\mu(P)I_{\rm bin}(\gamma d_P^2,w_P)$ over splitting parents, divided by $\sum_P\mu(P)H_b(w_P)$. Its half-time is read off without fitted parameters. Fig.~\ref{fig:mass_compare} compares prediction errors against the measured half-times. This test is sensitive to within-cell fluctuations and differences between splitting contrast and leading covariance eigenvector: it is not an exact alternative to the spectral law.

For all $60$ original Bethe instances at each of $\beta_J=1,1.5,2.5$ and $N=16$, and all $24$ corresponding instances at $N=20$, the mean residual declines when replacing mirror sectors by terminal cells. For example, at $\beta_J=1.5$ it falls from $0.1162$ to $0.0640$ ($N=16$) and from $0.0994$ to $0.0606$ ($N=20$). The depth caps and acceptance gates remain exactly as in Sec.~\ref{sec:levels}. For the prespecified $N=24$, $\beta_J=2.5$ instance analyzed in Fig.~\ref{fig:anatomy}, the residual decreases from $0.14242$ for two sectors to $0.01222$ for fourteen terminal cells (see Fig.~\ref{fig:anatomy_overlap}). Consequently, $96.7\%$ of the exact second overlap moment is represented by the terminal magnetizations in this \emph{particular} example.

For the $553$ available exact glassy-level half-times of Bethe and SK instances at $N=16,20$, the median $|\log(\gamma_{\rm pred}/\gamma_{1/2})|$ is $0.183$ for the spectral approximation and $0.086$ for the mass-aware centroid channel, with the latter closer to the measured time in $81.6\%$ of the comparisons. These measurements are correlated across levels and disorder conditions; the numbers describe this fixed archived campaign and do not independently validate rigid-centroid dynamics.


\section{Numerical methods}\label{app:methods}
\paragraph{Exact enumeration to $N=24$.} Configurations are written as $y=(y_a,y_b)$ with halves of $N/2$ spins, so that log-weights, field
terms $h_a\cdot y_a+h_b\cdot y_b$ and class labels are $2^{N/2}\times2^{N/2}$ arrays. A compiled kernel returns, for each field, the
row and column marginals (hence $\yh$) and the class masses in one pass ($0.13$\,s per field at $N=24$ on two cores); the covariance
$C_\gamma$ is assembled from three block products. Trajectories are generated endpoint first, $H_{\gamma_k}=\gamma_kY+B_{\gamma_k}$, on $25$ logarithmic grid points
$\gamma\in[0.02/N,5]$ plus $\gamma=0$. Endpoints $Y$ are sampled exactly and stratified over the finest classes: each class $k$ of mass
$\omega_k$ receives $n_k=\max(2,\mathrm{round}(n\omega_k))$ endpoints with weights $\omega_k/n_k$, which keeps small classes represented; $n=256$, $96$
and $48$ at $N=16$, $20$ and $24$, and $128$ for the instance of Fig.~\ref{fig:anatomy}. Mutual informations are
$\Hh(C_\ell)-\E\,\Hh(C_\ell\mid H_\gamma)$ with exact class posteriors and the stratification weights; half-times are interpolated in
$\log\gamma$, and their $68\%$ intervals are obtained by resampling paths ($60$ bootstrap replicas). Classes and predicted decision times alone, without
trajectories (Sec.~\ref{sec:procedure}), take about $0.1$\,s per instance at $N=16$ and $1$--$3$ minutes at $N=24$; the full analysis with
trajectories takes $2$\,s, $15$\,s and $5$ minutes at $N=16$, $20$ and $24$. Code: \texttt{code/exacttree.py}, \texttt{experiments/n1\_levels.py}, \texttt{experiments/n1\_figures.py}.

\paragraph{Extra overlap and unequal-mass checks.} The supplementary overlap/channel diagnostic script in \texttt{extra\_code/} reconstructs each Bethe disorder sample using the same \texttt{J\_of} and \texttt{build\_tree} functions as the published experiment. Its overlap residual uses exact Gibbs weights and full (mirror-symmetrized) class labels; its channel calculation uses Gauss--Hermite integration of Eq.~\eqref{eq:unbalanced_binary} and a numerical half-information root.  \texttt{check\_anatomy\_n24.py} reconstructs the prespecified $N=24$ example and exports its depth series.

\paragraph{Per-trajectory counting.} For Figs.~\ref{fig:partial}d--f and \ref{fig:families}c, d the posteriors are recomputed on $40$ logarithmic grid points of $\gamma$ ($128$
trajectories per instance at $N=16$, $64$ at $N=20$), and for
every trajectory the commitment time---the first grid value of $\gamma$ at which $p_\ell=p_{c_\ell(Y)}\ge0.9$---is recorded at each level $\ell$ at
which the trajectory's branch splits, $c_\ell(Y)\ne c_{\ell-1}(Y)$. Commitments within a factor $f$ in
$\gamma$ are merged into one event; the grid spacing is a factor $1.23$, below the merging factor. The mean number of events at $N=16$ stays between
$1.3$ and $1.8$ (default $1.5$) when the commitment threshold is varied between $0.8$ and $0.95$ and $f$ between $1.5$ and $3$. We also tried a label-free count, the number
of peaks of the per-trajectory localization rate $\gamma\operatorname{Tr}C_\gamma^2/N$; at these sizes single-trajectory traces fluctuate too much
for a reliable count, and we report only the label-based one. Panels a--c use the stored class posteriors of the instance of
Fig.~\ref{fig:anatomy}. Code: \texttt{experiments/n2\_paths.py}, \texttt{experiments/n3\_partial.py}.

\paragraph{Discretized generator.} For Fig.~\ref{fig:protocol} the drift $\kappa\yh(\kappa x)-cx$ is computed exactly on the configurations whose
prior weight exceeds $e^{-25}$ times the largest (discarded mass below $10^{-6}$). Euler steps are uniform in $t$, $t_k=k/K$, $k=0,\dots,K-1$; the
endpoint is $y=\operatorname{sign}(X_1)$. The tuned stiffness solves $S/\sinh S=\gamma_*$, so that $\gamma(1/2)=\gamma_*$, the geometric mean of the
predicted sector and first glassy decision times. The sampling-noise floor of the class error is the mean total variation of $4000$ exact samples
($20$ replicas). Code: \texttt{experiments/n2\_discretization.py}.

\paragraph{Larger systems ($N=32,64$).} For the six selected Bethe disorders, equilibrium libraries come from heat-bath parallel tempering
(28 temperatures, four independent histories, $238$--$465$ round trips per replica, random global signs added). Posteriors along the
paths are computed by sequential Monte Carlo (likelihood increments, resampling, Gibbs rejuvenation) on a $21$-point grid up to $\gamma=3.2$;
when independent populations disagree, a fixed-field parallel-tempering anchor is inserted without changing the path. $142$ of $144$ paths
meet the repeatability criterion and four audited field/time points agree with fresh references to total variation $0.004$--$0.008$. In the
sector window ($\gamma\le0.085$) the Nishimori check $\E[\yh\cdot Y-|\yh|^2]/N$ fluctuates with standard deviation $0.11$--$0.14$ over the $24$
paths and three disorders, with no consistent sign (pooled means $+0.07$ at $N=32$, $-0.04$ at $N=64$).

\paragraph{Learned denoisers.} One-hidden-layer networks
$\hat m_\theta(h,\gamma)=\tanh\!\big(V\tanh(Wh+c\log\gamma+b)+D\odot h+b'\big)$ are trained on the denoising loss with
$\log\gamma\sim\mathcal U[\log10^{-3},\log10]$ (Adam, one-cycle schedule, $1.2\times10^4$ steps of batch $512$), widths $1$--$32$, three seeds at
width $16$. Unit information is a plug-in estimate with $16$ equal-mass bins and the Miller--Madow correction on $4000$ test pairs per
$\gamma$; ablation importance is the increase of test error when a unit is replaced by its mean. The memorization study uses three hidden
layers of $256$ units trained on $64$, $1024$ or unlimited configurations of one instance.

\FloatBarrier
\section{Reproducibility}\label{app:repro}
{\sloppy All material is in the repository \url{https://github.com/mchertkov/SpinGlassHierarchy}. The folder \texttt{code} contains the
libraries: \texttt{exacttree.py} (exact enumeration, classes, endpoint-first trajectories and exact posteriors for $N\le24$) and \texttt{pidlib.py},
\texttt{pidlib\_lite.py} (sampling, sequential Monte Carlo and the positive control). The folder \texttt{experiments} contains the scripts that
generate the data and those that draw the figures; \texttt{results} contains the archived data; \texttt{paper} the source and figures of this paper;
and \texttt{notebooks} three executed notebooks: numerical checks of the theory, a complete analysis of one $N=16$ instance from scratch (classes,
predicted decision times, trajectories, partial commitments and information per level, in under a minute), and the regeneration of all figures.
The command \texttt{python scripts/make\_figures.py} redraws every figure from the archived data in a few minutes. The recomputation of the
data, in execution order, is documented in \texttt{scripts/run\_all\_experiments.sh}: the exact campaign (\texttt{n1\_run\_all.sh},
\texttt{n1\_run\_extension.sh}, \texttt{n2\_run.sh}, \texttt{n2\_discretization.py}, \texttt{n3\_partial.py}, \texttt{n4\_run\_sk.sh}; about
seven hours on two cores), the positive control (\texttt{e0\_caricature.py}, \texttt{s8\_validation.py}; minutes), and the sampled and learned
studies (\texttt{e5b\_*}, \texttt{e5c\_*}, \texttt{e5e\_*}, \texttt{e6\_*}, \texttt{s7\_*}; hours).\par}
\begin{center}\small
\begin{tabular}{@{}lll@{}}\toprule
Figure & Figure script (\texttt{experiments/}) & Data (\texttt{results/}) \\ \midrule
\ref{fig:schematic} & \texttt{n2\_schematic.py} & numbers of Fig.~\ref{fig:anatomy} \\
\ref{fig:anatomy} & \texttt{n1\_figures.py} & \texttt{N1/*anatomy*} \\
\ref{fig:partial} & \texttt{n2\_figures.py} & \texttt{N3}, \texttt{N2/paths\_*} \\
\ref{fig:temperature} & \texttt{n1\_figures.py} & \texttt{N1S} \\
\ref{fig:scan} & \texttt{n1\_figures.py} & \texttt{N1S}, \texttt{N1} \\
\ref{fig:law} & \texttt{n1\_figures.py} & \texttt{N1} \\
\ref{fig:families} & \texttt{n4\_families.py} & \texttt{N1}, \texttt{N1S}, \texttt{N2}, \texttt{N4} \\
\ref{fig:protocol} & \texttt{n2\_figures.py} & \texttt{N2/disc\_*} \\
\ref{fig:controls} & \texttt{e5e\_summarize.py} & \texttt{E5E}, \texttt{E5F}, \texttt{S8} \\
\ref{fig:larger} & \texttt{e5d\_analyze.py} & \texttt{E5B}--\texttt{E5D} \\
\ref{fig:learned} & \texttt{e6\_figures\_b.py} & \texttt{E6/E6B}, \texttt{S7} \\
\ref{fig:E0model}, \ref{fig:E0} & \texttt{fig\_E0\_model.py}, \texttt{fig\_E0.py} & \texttt{E0}, \texttt{S8} \\
\ref{fig:tools} & \texttt{e6\_figures\_a.py} & \texttt{E6/E6A} \\
\ref{fig:legacystatic} & \texttt{s7\_figure.py} & \texttt{S7}, \texttt{E5F} \\
\ref{fig:overlap_residual}, \ref{fig:mass_compare}, \ref{fig:anatomy_overlap} & \texttt{extra\_code/} & \texttt{extra\_results/} \\ \bottomrule
\end{tabular}\end{center}

\end{document}